\documentclass{article}

\usepackage{ifpdf}

\ifpdf
  \usepackage[pdftex]{color,graphicx}
  \usepackage[pdftex]{hyperref}
  \pdfpagewidth=\paperwidth
  \pdfpageheight=\paperheight

\else
  \usepackage[dvips]{epsfig}
  \DeclareGraphicsExtensions{.ps,.eps}
  \usepackage[dvips]{hyperref}
\fi  

\usepackage{amsmath,amssymb,amsthm,amsfonts}
\usepackage{triotex}
\usepackage{mathrsfs}

\theoremstyle{plain}
\newtheorem{theorem}{Theorem}

\theoremstyle{definition}

\newtheorem{example}[theorem]{Example}

\newcommand{\kw}[1]{\textsc{#1}}

\newcommand{\DClen}{\ell}

\newcommand{\LTLopletter}[1]{\mathrm{#1}}
\newcommand{\XLTL}{\LTLopletter{X}}
\newcommand{\FLTL}{\LTLopletter{F}}
\newcommand{\GLTL}{\LTLopletter{G}}
\newcommand{\ULTL}{\LTLopletter{U}}
\newcommand{\PLTL}{\LTLopletter{P}}
\newcommand{\OLTL}{\LTLopletter{O}}
\newcommand{\SLTL}{\LTLopletter{S}}
\newcommand{\ACTL}{\LTLopletter{A}}
\newcommand{\EXCTL}{\LTLopletter{E}}
\newcommand{\pp}{\mathsf{p}}
\newcommand{\qq}{\mathsf{q}}

\newcommand{\PP}{\mathsf{P}}

\newcommand{\eventin}{\;event\;in\;}
\newcommand{\completein}{\;complete\;in\;}
\newcommand{\Alwgran}[2]{\TRIOop{Alw}{#1}{#2}}

\newcommand{\hr}{\mathsf{hr}}
\newcommand{\lr}{\mathsf{lr}}
\newcommand{\wrr}{\mathsf{wr}}
\newcommand{\woner}{\mathsf{w1r}}
\newcommand{\hpr}{\mathsf{hpr}}
\newcommand{\lpr}{\mathsf{lpr}}
\newcommand{\wlr}{\mathsf{wlr}}
\newcommand{\slr}{\mathsf{slr}}
\newcommand{\pendh}{\mathsf{pendh}}
\newcommand{\noop}{\mathsf{noop}}
\newcommand{\nohr}{\mathsf{no\text{-}hr}}
\newcommand{\nolr}{\mathsf{no\text{-}lr}}
\newcommand{\hhr}{\mathsf{hhr}}
\newcommand{\rel}{\mathsf{rel}}
\newcommand{\free}{\mathsf{free}}
\newcommand{\occ}{\mathsf{occ}}
\newcommand{\glr}{\mathsf{glr}}
\newcommand{\ghr}{\mathsf{ghr}}
\newcommand{\wg}{\mathsf{wg}}
\newcommand{\too}{\mathsf{to}}

\newcommand{\distance}[1]{\mathrm{d}\!\left(#1\right)}

\newcommand{\statex}{\mathbf{x}}
\newcommand{\inputu}{\mathbf{u}}
\newcommand{\outputy}{\mathbf{y}}

\newcommand{\languageL}{\mathcal{L}}
\newcommand{\sdomain}{S}

\newcommand{\stvar}{\mathbf{s}}

\title{Modeling Time in Computing: \\ A Taxonomy and a Comparative Survey}

\author{Carlo A. Furia, Dino Mandrioli, \\ Angelo Morzenti, and Matteo Rossi}

\begin{document}

\maketitle

\begin{abstract}
The increasing relevance of areas such as real-time and embedded
systems, pervasive computing, hybrid systems control, and biological
and social systems modeling is bringing a growing attention to the
temporal aspects of computing, not only in the computer science
domain, but also in more traditional fields of engineering.

This article surveys various approaches to the formal modeling 
and analysis of the temporal features of computer-based systems, 
with a level of detail that is also suitable for nonspecialists. 
In doing so, it provides a unifying framework, rather than just 
a comprehensive list of formalisms.

The article first lays out some key dimensions along which the 
various formalisms can be evaluated and compared. Then, a significant 
sample of formalisms for time modeling in computing are presented 
and discussed according to these dimensions. The adopted perspective 
is, to some extent, historical, going from ``traditional''
models and formalisms to more modern ones.
\end{abstract}

\newpage

\tableofcontents

\newpage

\section{Introduction} \label{sec:introduction}
In many fields of science and engineering, the term \emph{dynamics} is intrinsically bound to a notion of time. In fact, in classical physics a mathematical model of a dynamical system most often consists of a set of equations that state a relation between a \emph{time variable} and other quantities characterizing the system, often referred to as system \emph{state}.

In the theory of computation, conversely, the notion of time does not always play a major role. At the root of the theory, a problem is formalized as a \emph{function} from some input domain to an output range. An algorithm is a process aimed at computing the value of the function; in this process, dynamic aspects are usually abstracted away, since the only concern is the result produced.

Timing aspects, however, are quite relevant in computing too, for many reasons; let us recall some of them by adopting a somewhat historical perspective.

\begin{itemize}
\item First, \emph{hardware} design leads down to electronic devices where the physical world of circuits comes back into play, for instance when the designer must verify that the sequence of logical gate switches that is necessary to execute an instruction can be completed within a clock's tick. The time models adopted here are borrowed from physics and electronics, and range from differential equations on continuous time for modeling devices and circuits, to discrete time (coupled with discrete mathematics) for describing logical gates and digital circuits.

\item When the level of description changes from hardware to software, physical time is progressively disregarded in favor of more ``coarse-grained'' views of time, where a time unit represents a computational step, possibly in a high-level programming language; or it is even completely abstracted away when adopting a purely functional view of software, as a mapping from some input to the computed output. In this framework, \emph{computational complexity} theory was developed as a natural complement of computability theory: it was soon apparent that knowing an algorithm to solve a problem is not enough if the execution of such an algorithm takes an unaffordable amount of time. As a consequence, models of abstract machines have been developed or refined so as to measure the time needed for their operations. Then, such an abstract notion of time measure (typically the number of elementary computation steps) could be mapped easily to physical time.

\item The advent of \emph{parallel} processing mandated a further investigation of timing issues in the theory of computing. To coordinate appropriately the various concurrent activities, in fact, it is necessary to take into account their temporal evolution. Not by chance the term synchronization derives from the two Greek words $\sigma \upsilon \nu$ (meaning ``together'') and $\chi \rho o \nu o \sigma$ (meaning ``time'').

\item In relatively recent times the advent of novel methods for the design and verification of \emph{real-time} systems also requires the inclusion of the environment with which the computer interacts in the models under analysis. Therefore the various activities are, in general, not fully synchronized, that is, it is impossible to delay indefinitely one activity while waiting for another one to come alive. Significant classes of systems that possess real-time features are, among others, social organizations (in a broad sense), and distributed and embedded systems. For instance, in a plant control system, the control apparatus must react to the stimuli coming from the plant at a pace that is mandated by the dynamics of the plant. Hence physical time, which was progressively abstracted away, once again plays a prominent role.
\end{itemize}

\noindent As a consequence, some type of time modeling is necessary in the theory of computing as well as in any discipline that involves dynamics. Unlike other fields of science and engineering, however, time modeling in computing is far from exhibiting a unitary and comprehensive framework that would be suitable in a general way for most needs of system analysis: this is probably due to the fact that the issue of time modeling arose in different fields, in different circumstances, and was often attacked in a fairly \emph{ad hoc} manner.

In this article we survey various approaches that have been proposed to tackle the issue of time modeling in computing. Rather than pursuing an exhaustive list of formalisms, our main goal is to provide a unifying framework so that the various models can be put in perspective, compared, evaluated, and possibly adapted to the peculiar needs of specific application fields. In this respect, we selected the notations among those that are most prominent in the scientific literature, both as basic research targets and as useful modeling tools in applications. We also aimed at providing suitable ``coverage'' of the most important features that arise in time modeling. We tried to keep our exposition at a level palatable for the nonspecialist who wishes to gain an overall but not superficial understanding of the issue. Also, although the main goal of time modeling is certainly to use it in the practice of system design, we focus on the conceptual aspects of the problem (what can and cannot be done with a given model; how easy it is to derive properties, etc.) rather than on practical ``recipes'' of how to apply a formal language in specific projects. The presentation is accompanied by many examples from different domains; most of them are inspired by embedded systems concepts, others, however, show that the same concepts apply as well to a wider class of systems such as biological and social ones.

We deliberately excluded from our survey time modeling approaches based on stochastic formalisms. This sector is certainly important and very relevant for several applications, and it has recently received increasing attention from the research community (e.g., \cite{RK+04,DK05}). In fact, most of the formal notations presented in this survey have some variants that include stochastic or probabilistic features. However, including such variants in our presentation would have also required us to present the additional mathematical notions and tools needed to tackle stochastic processes. These are largely different from the notions discussed in the article, which aim at gaining ``certainty'' (e.g., ``the system will not crash under any circumstances'') rather than a ``measure of uncertainty'' (e.g., ``the system will crash with probability $10^{-3}$'') as happens with probabilistic approaches. Thus, including stochastic formalisms would have weakened the focus of the article and made it excessively long.

The first part of this article introduces an informal reference framework within which the various formalisms can be explained and evaluated. First, Section \ref{sec:metamodel} presents the notion of language, and gives a coarse categorization of formalisms; then, Section \ref{sec:dimensions} proposes a collection of ``dimensions'' along which the various modeling approaches can be classified.

The second part of the article is the actual survey of time modeling formalisms. We do not aim at exhaustiveness; rather, we focus on several relevant formalisms, those that better exemplify the various approaches found in the literature, and analyze them through the dimensions introduced in Section \ref{sec:dimensions}. We complement the exposition, however, with an extensive set of bibliographic references. In the survey, we follow a rather historical ordering: Section \ref{sec:historical} summarizes the most traditional ways of taking care of timing aspects in computing, whereas Section \ref{sec:modernmodels} is devoted to the more recent proposals, often motivated by the needs of new, critical, real-time applications. Finally, Section \ref{sec:discussion} contains some concluding remarks.



\section{Languages and Interpretations} \label{sec:metamodel}
When studying the different ways in which time has been represented in the literature, and the associated properties, two aspects must be considered: the language used to describe time and the way in which the language is interpreted.\footnote{Such interpretations are referred to in mathematical 
logic as the \emph{models} of $\phi$; in this article we will 
in general avoid this terminology, as it might generate some 
confusion with the different notion of a model as ``description 
of a system''.} Let us illustrate this point in some detail.

A \emph{language} (in the broad sense of the term) is the device that we employ to describe anything of interest (an object, a function, a system, a property, a feature, etc.). Whenever we write a ``sentence'' in a language (any language), a \emph{meaning} is also attached to that sentence. Depending on the extent to which mathematical concepts are used to associate a sentence with its meaning, a language can be informal (no elements of the language are associated with mathematical concepts), semiformal (some are, but not all), or formal (everything is).

More precisely, given a sentence $\phi$ written in some language $\languageL$, 
we \emph{can} assign it a variety of \emph{interpretations}; we then 
define the meaning of $\phi$ by establishing which ones, 
among all possible interpretations, are those that are \emph{actually 
associated} with it (in other words, by deciding which interpretations 
are ``meaningful'', and which ones are not); we say that 
an interpretation \emph{satisfies} a sentence with which it is associated, 
or, dually, that the sentence \emph{expresses} its associated interpretations. 
In the rest of this article, we will sometimes refer to the language 
as the \emph{syntax} used to describe a sentence, as opposed to 
the interpretations that the latter expresses, which constitute 
its \emph{semantics}.

In this survey we mainly deal with languages that have the distinguishing feature of including a notion of time. Then, the interpretations associated with sentences in these languages include a notion of \emph{temporal evolution} of elements; that is, they define what value is associated with an element at a certain time instant. As a consequence, we refer to the possible interpretations of sentences in timed languages as \emph{behaviors}. In fact, the semantics of every formal language that has a notion of time is defined through some idea of ``behavior'' (or trace): infinite words for linear temporal logic \cite{Eme90}, timed words for timed automata \cite{AD94}, sequences of markings for Petri nets \cite{Pet81}, and so on.

For example, a behavior of a system is a mapping $b: \timedomain \rightarrow \sdomain$, where $\timedomain$ is a temporal domain and $\sdomain$ is a state space; the behavior represents the system's state (i.e., the value of its elements) in the various time instants of $\timedomain$.

Let us consider a language $\languageL$ and a sentence $\phi$ written 
in $\languageL$. The nature of $\phi$ depends on $\languageL$; for example 
it could be a particular kind of graph if $\languageL$ is some type 
of automaton (a Statechart, a Petri net, etc.), a logic formula 
if $\languageL$ is some sort of logic, and so on. Given a behavior $b$ in 
the system model, we write $b \models \phi$ to indicate that $b$
satisfies $\phi$, that is, it is one of the behaviors expressed 
by the sentence. The satisfaction relation $\models$ is not general, that is, it 
is language-dependent (it is, in fact, $\models_{\languageL}$, but we omit the subscript 
for conciseness), and is part of the definition of the language.

\begin{figure}[htb!]
	 \centering
	 \includegraphics{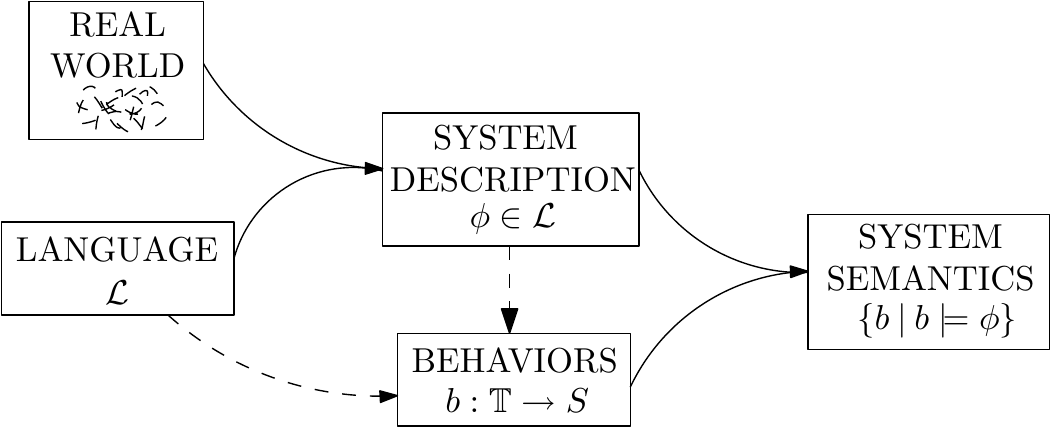}
	 \caption{Behaviors, language, system descriptions, world.}
	 \label{fig:worlds}
\end{figure}

Figure \ref{fig:worlds} depicts informally the relations among
behaviors, language, system descriptions, real world, and
semantics. Solid arrows denote that the entities they point to are
obtained by combining elements of entities they originate from; for
instance, a system description consists of formalizations of (parts
of) the real world through sentences in some language. Dashed arrows,
on the other hand, denote indirect influences; for example, the
features of a language can suggest the adoption of certain behavioral
structures. Finally, the semantics of a system is given by the set of
all behaviors $b$ satisfying system description $\Phi$. These
relations will become clearer in the following examples.

\begin{example}[Continuous, Scalar Linear Dynamic System]
Suppose $\languageL$ is the language of differential equations used to describe traditional 
linear dynamic systems. With such a language we might model, 
for example, the simple RC circuit of Figure \ref{fig:RCcircuit}.
In this case, the sentence $\phi$ that describes the system could be $\dot{q} = - \frac{1}{R C} q$
(where $q$ is the charge of the capacitor); then, a behavior $b$
that satisfies $\phi$ (i.e., such that $b \models \phi$)
is $b(t) = C_0 e^{-t/RC}$, where $C_0$ is the initial charge of the capacitor, at the time 
when the circuit is closed (conventionally assumed to be 0).
\begin{figure}[htb!]
	 \centering
	 \includegraphics{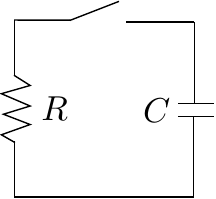}
	 \caption{An example of sentence in graphical language describing electric circuits.}
	 \label{fig:RCcircuit}
\end{figure}
\end{example}

To conclude this section, let us present a widely used categorization 
of languages that, while neither sharp nor precise, nevertheless 
quickly conveys some important features of a language.

Languages are often separated into two broad classes: \emph{operational} 
languages and \emph{descriptive} languages \cite{GJM02}.

Operational languages are well-suited to describe the \emph{evolution} 
of a system starting from some initial state. Common examples 
of operational languages are the differential equations used 
to describe dynamic systems in control theory (see Section \ref{sec:dynamicalsys}), 
automata-based formalisms (finite-state automata, Turing machines, 
timed automata, which are described in Sections \ref{sec:swview} and \ref{sec:synchronous}) 
and Petri nets (which are presented in Section \ref{sec:petrinets}). Operational 
languages are usually based on the key concepts of \emph{state} 
and \emph{transition} (or \emph{event}), so that a system is modeled 
as evolving from a state to the next one when a certain transition/event 
occurs. For example, an operational description of the dynamics 
of a digital safe could be the following: 

\begin{example}[Safe, operational formulation] \label{ex:operational}
``When the last digit of the correct security code is entered, the safe opens; 
if the safe remains open for three minutes, it automatically closes.''
\end{example}

Descriptive languages, instead, are better suited to describing 
the \emph{properties} (static or dynamic) that the system must satisfy. 
Classic examples of descriptive languages are logic-based and algebra-based formalisms, 
such as those presented in Section \ref{sec:descriptive}. An example of descriptive 
formulation of the properties of a safe is the following: 

\begin{example}[Safe, descriptive formulation] \label{ex:descriptive}
``The safe is open if and only if the correct security code has been entered no 
more than three minutes ago.'' 
\end{example}

As mentioned above, the distinction between operational and descriptive 
languages is not as sharp as it sounds, for the following reasons. 
First, it is possible to use languages that are 
operational to describe system properties (e.g., \cite{AD94} 
used timed automata to represent both the system and its properties 
to be verified through model checking), and languages that are 
descriptive to represent the system evolution with 
state/event concepts \cite{GM01} (in fact, the dynamics of Example \ref{ex:operational} 
can be represented using a logic language, while the property 
of Example \ref{ex:descriptive} can be formalized through an automata-based language). 
In addition, it is common to use a combination of operational 
and descriptive formalisms to model and analyze systems in a 
so-called \emph{dual-language approach}. In this dual approach, 
an operational language is used to represent the dynamics of 
the system (i.e., its evolution), while its requirements (i.e., the 
properties that it must satisfy, and which one would like to 
verify in a formal manner) are expressed in a descriptive language. 
Model checking techniques \cite{CGP00,HNSY94} and the combination 
of Petri nets with the TRIO temporal logic \cite{FMM94} are examples 
of the dual language approach.


\section{Dimensions of the Time Modeling Problem} \label{sec:dimensions}
When describing the modeling of time several distinctive issues need
to be considered. These constitute the ``dimensions'' of the problem
from the perspective of this article. They will help the analysis of how
time is modeled in the literature, which is carried out in Sections \ref{sec:historical} and \ref{sec:modernmodels}.

Some of the dimensions proposed here are indicative of issues that are pervasive in the modeling of time in the literature (e.g., using discrete vs.~continuous time domains); others shed more light on subtle aspects of some formalisms. We believe that the systematic, though not exhaustive, analysis of the formalisms surveyed in Sections \ref{sec:historical} and \ref{sec:modernmodels} against the dimensions proposed below should not only provide the reader with an overall comparative assessment of the formalisms described in this article, but also help her build her own evaluation of other present and future formalisms in the literature.

\subsection{Discrete vs.~Dense Time Domains} \label{sec:discrete}
A first natural categorization of the formalisms dealing with 
time-dependent systems and the adopted time model is whether 
such a model is a discrete or dense set.

A discrete set consists of isolated points, whereas a dense set (ordered by ``$<$'')
is such that for every two points $t_1, t_2$, with $t_1 < t_2$, there is always another 
point $t_3$ in between, that is, $t_1 < t_3 < t_2$. In the scientific literature and applications, 
the most widely adopted discrete time models are natural and 
integer numbers --- herewith denoted as $\naturals$ and $\integers$, respectively 
--- whereas the typical dense models are rational and real numbers 
--- herewith denoted as $\rationals$ and $\reals$, respectively. For instance, 
differential equations are normally stated with respect to real 
variable domains, whereas difference equations are 
defined on integers. Computing devices are formalized through 
discrete models when their behavior is paced by a clock, so that 
it is natural to measure time by counting clock ticks, or when 
they deal with (i.e., measure, compute, or display) values in 
discrete domains.

Besides the above well-known premises, however, a few more accurate 
distinctions are useful to better evaluate and compare the many 
formalisms available in the literature and those that will be 
proposed in the future.

\subsubsection*{Continuous vs.~Noncontinuous Time Models}
Normally in mathematics, continuous time models (i.e., those in which the temporal domain is a dense set such that every nonempty set with an upper bound has a least upper bound) such as real 
numbers are preferred to other dense domains such 
as the rationals, thanks to their completeness/closure with respect 
to all common operations (otherwise, referring to $\sqrt{2}$ or $\pi$
would be cumbersome). Instead, normal numerical algorithms deal 
with rational numbers since they can approximate real numbers 
--- which cannot be represented by a finite sequence of digits 
--- up to any predefined error. There are cases, however, where 
the two sets exhibit a substantial difference. For instance, 
assume that a system is composed of two devices whose clocks $c_1$ 
and $c_2$ are incommensurable (i.e., such that there are no 
integer numbers $n, m$ such that $n c_1 = m c_2$). In such a case, if one 
wants to ``unify'' the system model, $\rationals$ is not a suitable temporal domain.
Also, there are some sophisticated time analysis algorithms 
that impose the restriction that the time domain is $\rationals$ but 
not $\reals$. We refer to one such algorithm when discussing Petri nets in Section \ref{sec:petrinets}.

\subsubsection*{Finite or Bounded Time Models}
Normal system modeling assumes behaviors that may proceed indefinitely 
in the future (and maybe in the past), so that it is natural 
to model time as an unbounded set. There are significant cases, 
however, where all relevant system behaviors can be \emph{a priori} 
enclosed within a bounded ``time window''. For instance, braking 
a car to a full stop requires at most a few seconds; thus, if 
we want to model and analyze the behavior of an anti-lock braking 
system there is no loss of generality if we assume as a temporal domain, say,
the real range $[0 \ldots 60\text{secs}]$. In many cases this 
restriction highly simplifies several analyses and/or simulation 
algorithms. In other cases the system under consideration is 
periodic; thus, knowing its behaviors during a full period provides 
enough information to determine its relevant properties over the whole time axis.

\subsubsection*{Hybrid Systems}
In this article, by hybrid system model we mean a model that uses 
both discrete and dense domains. There are several circumstances 
when this may occur, mainly but not exclusively related 
to the problem of integrating heterogeneous components: for 
instance, monitoring and controlling a continuous process by 
means of a digital device. 

\begin{itemize}
\item A system (component) with a discrete --- possibly finite --- set 
of states is modeled as evolving in a dense time domain. In such 
a case its behavior is graphically described as a \emph{square wave} form (see Figure \ref{fig:squarewave})
and its state can be formalized as a piecewise constant function of time, as shown in Figure \ref{fig:squarewave}.

\begin{figure}[htb!]
	 \centering
	 \includegraphics{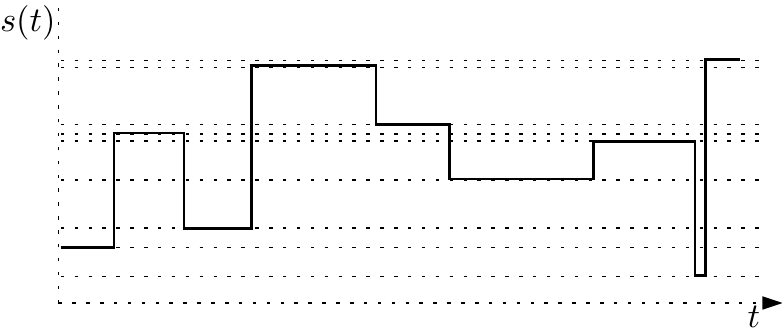}
	 \caption{A square-wave form over dense time.}
	 \label{fig:squarewave}
\end{figure}

\item In a fairly symmetric way, a continuous behavior can be sampled 
at regular intervals, as exemplified in Figure \ref{fig:sampledsinus}. 

\begin{figure}[htb!]
	 \centering
	 \includegraphics[scale=0.7]{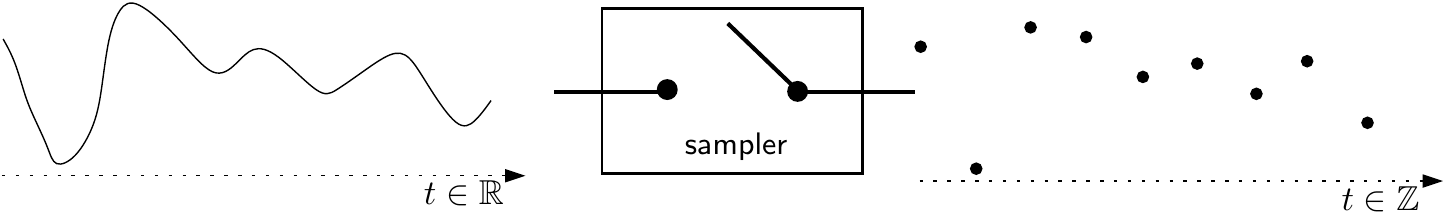}
	 \caption{A sampled continuous behavior.}
	 \label{fig:sampledsinus}
\end{figure}

\item A more sophisticated, but fairly common, case of hybridness may
  arise when a model evolves through a discrete sequence of ``steps''
  while other, independent, variables evolve taking their values in
  nondiscrete domains, for instance, finite state automata augmented
  with dense-timed clock variables.  We see examples of such
  models in Section \ref{sec:operational}, in which timed and hybrid
  automata are discussed.
\end{itemize}

\subsubsection*{Time Granularity}
In some sense, time granularity can be seen as a special case of
hybridness. We say that two system components have different time
granularities when their ``natural time scales'' differ, possibly by
orders of magnitude. This, again, is quite frequent when we pair a
process that evolves in the order of seconds or minutes, or even days
or months (such as a chemical plant, or a hydroelectric power plant)
with a controller based on digital electronic devices.  In principle,
if we assume a unique continuous time model, say the reals, the
problem is reduced to a, possibly cumbersome, change of time
unit.\footnote{Notice, however, that in very special cases the
  different time units could be incommensurable. In fact, even if in
  practice this circumstance may seldom arise, after all the two main
  units offered by nature, the day and the year, are incommensurable.}

However, if discrete domains are adopted, subtle semantic issues
related to the approximation of the coarser time unit may
arise. Consider, for instance, the sentences ``every month, if an
employee works, then she gets her salary'' and ``whenever an employee
is assigned a job, this job should be completed within three
days''. They may both be part of the same specification of an office
system, so an admissible behavior for the office system must
satisfy both sentences. It would be natural to assume a discrete
temporal domain in which the time unit is the day, which is of finer
granularity than the month. However, it is clear that stating that
``every month, if an employee works, then she gets her salary'' is not
the same as ``every day, if an employee works, then she gets her
salary''. In fact, in general, working for one month means that one
works for 22 days of the month, whereas getting a monthly salary means
that there is one day when one gets the salary for the month. Hence, a
simple change in the time unit (from months to days) in this case does
not achieve the desired effect.

As a further example, suppose that the sentence ``this job has to be
finished within 3 days from now'' is stated at 4 PM of a given day:
should this be interpreted as ``this job has to be finished within $3
\times 24 \times 60 \times 60$ seconds from now'', or ``this job has
to be finished before midnight of the third day after today''? Both
interpretations may be adopted depending on the context of the claim.

An approach that deals rigorously with different time granularities is
presented in Section \ref{sec:temporallogics} when discussing
temporal logics.

\subsection{Ordering vs.~Metric} \label{sec:ordering}
Another central issue is whether a formalism permits 
the expression of metric constraints on time, or, equivalently, 
of constraints that exploit the metric structure of the underlying 
time model (if it has any).

A domain (a time domain, in our case) has a \emph{metric} when a 
notion of \emph{distance} is defined on it (that is, when a nonnegative \emph{measure} $\distance{t_1, t_2} \geq 0$ is associated with any two points $t_1, t_2$ of the domain).

As mentioned above, typical choices for the time domain are the usual
discrete and dense numeric sets, that is $\naturals, \integers,
\rationals, \reals$.  All these domains have a ``natural'' metric
defined on them, which corresponds to simply taking the
distance\footnote{Technically, this is called the Euclidean distance.}
between two points: $\distance{t_1, t_2} = |t_1 - t_2 |$.\footnote{We
  focus our attention here on temporal domains $\timedomain$ that are
  totally ordered; although partially-ordered sets may be considered
  as time domains (see Section \ref{sec:concurrency}), they have not
  attracted as much research activity as totally ordered domains.}

Notice, however, that although all common choices for the time domains
possess a metric, we focus on whether the \emph{language} in which the
system is described permits descriptions using the same form of metric information as that embedded in the underlying time domain. For
instance, some languages allow the user to state that an event $\pp$
(e.g., ``push button'') must temporally precede another event $\qq$
(e.g., ``take picture''), but do not include constructs to specify how
long it takes between the occurrence of $\pp$ and that of $\qq$; hence,
they cannot distinguish the case in which the delay between $\pp$ and
$\qq$ is 1 time unit from the case in which the delay is 100 time
units. Thus, whenever the language does not allow users to state
metric constraints, it is possible to express only information about
the relative \emph{ordering} of phenomena (``$\qq$ occurs after
$\pp$''), but not about their distance (``$\qq$ occurs 100 time units
after $\pp$''). In this case, we say that the language has a purely
\emph{qualitative} notion of time, as opposed to allowing
\emph{quantitative} constraints, which are expressible with metric
languages.

Parallel systems have been defined \cite{Wir77} as those where the
correctness of the computation depends only on the \emph{relative
  ordering} of computational steps, irrespective of the absolute
distance between them. Reactive systems can often be modeled as parallel
systems, where the system evolves concurrently with the
environment. Therefore, for the formal description of such systems a
purely qualitative language is sufficient. On the contrary, real-time
systems are those whose correctness depends on the \emph{time
  distance} among events as well. Hence, a complete description of such systems
requires a language in which metric constraints can be expressed. In this
vein, the research in the field of formal languages for system
description has evolved from dealing with purely qualitative models to
the more difficult task of providing the user with the possibility of
expressing and reasoning about metric constraints.

For instance, consider the two sequences $b_1, b_2$ of events
$\pp,\qq$, where exactly one event per time step occurs:
\begin{itemize}
  \item $b_1 = \pp \qq \pp \qq \pp \qq \cdots$
  \item $b_2 = \pp \pp \qq \qq \pp \pp \qq \qq \cdots$
\end{itemize}
$b_1$ and $b_2$ share all the qualitative properties expressible
without using any metric operator. For instance ``every occurrence of
$\pp$ is eventually followed by an occurrence of $\qq$'' is an example
of qualitative property that holds for both behaviors, whereas ``$\pp$
occurs in every instant'' is another qualitative property, that instead is false for both behaviors. If referring to metric properties is
allowed, one can instead discriminate between $b_1$ and $b_2$, for
example through the property ``every occurrence of $\qq$ is followed
by another occurrence of $\qq$ after two time steps'', which holds for
$b_1$ but not for $b_2$.

Some authors have introduced a notion of equivalence between 
behaviors that captures the properties expressed by qualitative 
formulas. In particular Lamport \cite{Lam83} first proposed the notion 
of \emph{invariance under stuttering}. Whenever a (discrete time) 
behavior $b_3$ can be obtained from another behavior $b_4$ by adding and 
removing ``stuttering steps'' (i.e., pairs of identical states 
on adjacent time steps), we say that $b_3$ and $b_4$ are stutter-equivalent. 
For instance, behaviors $b_1$ and $b_2$ outlined above are stutter-equivalent. 
Then, the equivalence classes induced by this equivalence relation 
precisely identify classes of properties that share identical 
qualitative properties. Note that stutter invariance is defined 
for discrete time models only.

\subsection{Linear vs.~Branching Time Models} \label{sec:linear}
\begin{figure}[htb!]
	 \centering
	 \includegraphics[scale=1.2]{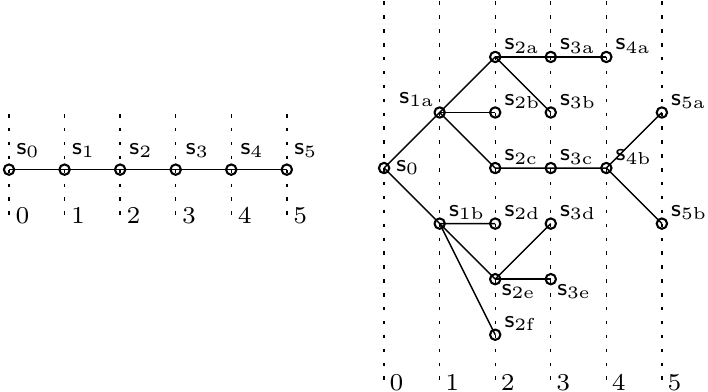}
	 \caption{A linear and a branching time model.}
	 \label{fig:linear-branching}
\end{figure}

The terms \emph{linear} and \emph{branching} refer to the structures 
on which a formal language is interpreted: \emph{linear}-time formalisms 
are interpreted over \emph{linear} sequences of states, whereas \emph{branching}-time 
formalisms are interpreted over \emph{trees} of states. In other 
words, each description of a system adopting a linear notion 
of time refers to (a set of) linear behaviors, where the future 
evolution from a given state at a given time is always unique. 
Conversely, a branching-time interpretation refers to behaviors 
that are structured in trees, where each ``present state'' may 
evolve into different ``possible futures''. For instance, assuming 
discrete time, Figure \ref{fig:linear-branching} pictures a linear sequence of states 
and a tree of states over six time instants.

A linear behavior is a special case of a tree. Conversely, 
a tree might be thought of as a set of linear behaviors that share 
common prefixes (i.e., that are prefix-closed); this notion is 
captured formally by the notion of \emph{fusion closure} \cite{AH92}. 
Thus, linear and branching models can be put on a common ground 
and compared. This has been done extensively in the literature.

Linear or branching semantic structures are then matched, in the
formal languages, by corresponding syntactic elements that allow us
to express properties about the specific features of the
interpretation. This applies in principle to all formal languages,
but it has been the focus of logic languages especially, and temporal
logics in particular. Thus, a linear-time temporal logic is
interpreted over linear structures, and is capable of expressing
properties of behaviors with unique futures, such as ``if event $\pp$
happens, then event $\qq$ will happen eventually in the future''. On
the other hand, branching-time temporal logics are interpreted over
tree structures and allow users to state properties of branching
futures, such as ``if event $\pp$ happens at some time $t$, then there
is some possible future where event $\qq$ holds''. We discuss this in
greater depth in our consideration of temporal logics (Section
\ref{sec:temporallogics}); for general reference we cite the classic
works by Lamport \cite{Lam80}, Emerson and Halpern \cite{EH86},
Emerson \cite{Eme90}, and Alur and Henzinger \cite{AH92} --- the last
focusing on real-time models.

Finally, we mention that it is possible to have semantic 
structures that are also \emph{branching in the past} \cite{Koy92}, which is where 
different pasts merge into a single present. However, in practice, 
branching-in-the-past models are relatively rare, so we will 
not deal with them in the remainder of this article.

\subsubsection*{Determinism vs. Nondeterminism}
Linear and branching time are features of the languages and structures on which they are interpreted, whereas the notions of
determinism and nondeterminism are attributes of the \emph{systems}
being modeled or analyzed. More precisely, let us consider systems
where a notion of \emph{input} is defined: one such system evolves
over time by reading its input and changing the current state
accordingly. Whenever the future state of the system is
\emph{uniquely} determined by its current state and input values, then
we call the system \emph{deterministic}. For instance, a light button
is a deterministic system where pressing the button (input) when the
light is in state \emph{off} yields the unique possible future state
of light \emph{on}. Notice that, for a given input sequence, the
behavior of a deterministic system is uniquely determined by its
initial state. Conversely, systems that can evolve to different future
states from the same present state and the same input by making
arbitrary ``choices'' are called \emph{nondeterministic}.  For
example, a resource arbiter might be a nondeterministic system that
responds to two requests happening at the same time by ``choosing''
arbitrarily to whom to grant the resource first.  The Ada programming
language \cite{BB94} embeds such a nondeterminism in its syntax and
semantics.

In linear-time models the future of any instant is unique, whereas in
branching-time models each instant branches into different possible
futures; then, there is a natural coupling between deterministic systems and linear models on one side, and on the other side
nondeterministic systems and branching models, where all possible
``choices'' are mapped at some time to branches in the tree. Often,
however, linear-time models are still preferred even for
nondeterministic systems for their intuitiveness and simplicity. In
the discussion of Petri nets (Section \ref{sec:petrinets}) we see
an example of linear time domains expressing the semantics of
nondeterministic formalisms.

\subsection{Implicit vs.~Explicit Time Reference} \label{sec:implicit}
Some languages allow, or impose on, the user to make explicit 
reference to temporal items (attributes or entities of ``type time''),
whereas other formalisms, though enabling reasoning about 
temporal system properties, leave all or some references to time-related 
properties (occurrences of events, durations of states or actions) 
implicit in the adopted notation.

To illustrate, at one extreme consider pure first-order 
predicate calculus to specify system behavior and its properties. 
In such a case we could use explicit terms ranging over the 
time domain and build any formula, possibly with suitable quantifiers, 
involving such terms. We could then could express properties such 
as ``if event $\pp$ occurs at instant $t$, then $\qq$ occurs at some 
instant $t'$ no later than $k$ time units after $t$''. At 
the other extreme, classic temporal logic \cite{Kam68}, despite its name, 
does not offer users the possibility to explicitly mention any 
temporal quantity in its formulas, but aims at expressing temporal 
properties by referring to an implicit ``current time'' and to 
the ordering of events with respect to it; for example, it has 
operators through which it is possible to represent properties 
such as ``if event $\pp$ occurs now, then sometime in the future 
event $\qq$ will follow''.

Several formalisms adopt a kind of intermediate approach.  For
instance, many types of abstract machines allow the user to specify
explicitly, say, the duration of an activity with implicit reference
to its starting time (e.g., Statecharts, discussed in Section
\ref{sec:synchronous}, and Petri nets, discussed in Section
\ref{sec:petrinets}). Similarly, some languages inspired by temporal
logic (e.g., MTL, presented in Section \ref{sec:temporallogics}) keep
its basic approach of referring any formula to an implicit current
instant (the \emph{now} time), but allow the user to explicitly express
a time distance with respect to it. Typical examples of such formulas
may express properties such as ``if event $\pp$ occurs now then event
$\qq$ will follow in the future within $t$ time units''.

In general, using implicit reference to time instants --- in particular 
the use of an implicit \emph{now} --- is quite natural and allows 
for compact formalizations when modeling and expressing properties 
of so-called ``time-invariant systems'', which are the majority 
of real-life systems. In most cases, in fact, the system behavior 
is the same if the initial state of and the input supplied to 
the system are the same, even if the same computation occurs 
at different times. Hence the resulting behaviors are simply 
a temporal translation of one another, and in such cases, explicitly expressing where the \emph{now} is located in the time 
axis is superfluous.

\subsection{The Time Advancement Problem} \label{sec:timeadvancement}
The problem of time advancement arises whenever the model of 
a timed system exhibits behaviors that do not progress past some 
instant. Such behaviors do not correspond to any physical 
``real'' phenomena; they may be the consequence of some incompleteness 
and inconsistency in the formalization of the system, and thus must 
be ruled out. 

The simplest manifestation of the time advancement problem arises 
with models that allow transitions to occur in a null time. For 
instance, several automata-based formalisms such as Statecharts 
and timed versions of Petri nets adopt this abstract notion (see 
Section \ref{sec:synchronous}). Although a truly instantaneous action is physically 
unfeasible, it is nonetheless a very useful abstraction for events 
that take an amount of time which is negligible with respect to 
the overall dynamics of the system \cite{BB06}. For example, pushing a button is an action whose actual duration can usually be ignored and 
thus can be represented abstractly as a zero-time event. When 
zero-time transitions are allowed, an infinite number of such 
transitions may accumulate in an arbitrarily small interval to the left
of a given time instant, thus modeling a fictitious infinite 
computation where time does not advance at all. Behaviors where 
time does not advance are usually called \emph{Zeno} behaviors, 
from the ancient philosopher Zeno of Elea\footnote{Circa 490--425 BC.}
and his paradoxes on time advancement (the term was coined 
by Abadi and Lamport \cite{AL94}). Notice that, from a rigorous point 
of view, even the notion of behavior as a function --- whose domain is
time and whose range is system state --- is ill-defined if zero-time transitions 
are allowed, since the consequences of a transition that takes 
zero time to occur is that the system is both at the source state 
and at the target state of the transition in the same instant.

Even if actions (i.e., state transformations) are noninstantaneous, 
it is still possible for Zeno behaviors to occur if time advances 
only by \emph{arbitrarily small} amounts. Consider, for instance, 
a system that delivers an unbounded sequence of events $\pp_k$, for $k \in \naturals$;
each event $\pp_k$ happens exactly $t_k$ time units after the previous one 
(i.e., $\pp_{k-1}$). If the sum of the relative times (that is, the sum $\Sigma_k t_k$
of the time distances between consecutive events) converges to 
a finite limit $t$, then the absolute time never surpasses $t$; in other 
words, \emph{time stops} at $t$, while an infinite number of events 
occur between any $t_k$ and $t$. Such behaviors allow an infinite number 
of events to occur within a finite time.

Even when we consider continuous-valued time-dependent functions 
of time that vary smoothly, we may encounter Zeno behaviors. 
Take, for instance, the real-valued function of time $b(t) = \exp{(-1/t^2)} \sin(1/t)$; $b(t)$
is very smooth, as it possesses continuous derivatives of all 
orders. Nonetheless, its sign changes 
an infinite number of times in any interval containing the origin; 
therefore a natural notion such as ``the next instant at which 
the sign of $b$ changes'' is not defined at time 0, and, consequently, 
we cannot describe the system by relating its behavior to such 
--- otherwise well-defined --- notions. Indeed, as explained
precisely in Section \ref{sec:descriptive} when discussing temporal logics, non-Zenoness can be mathematically 
characterized by the notion of \emph{analyticity}, which is even 
stronger than infinite derivability.

The following remarks are to some extent related to the problem 
of time advancement, and might help a deeper understanding thereof. 

\begin{itemize}

\item Some formal systems possess ``Zeno-like'' behaviors, where the
  distance between consecutive events gets indefinitely smaller, even
  if time progresses (these behaviors have been called ``Berkeley'' in
  \cite{FPR08}, from the philosopher George
  Berkeley\footnote{Kilkenny, 1685--Oxford, 1753.} and his
  investigations arguing against the notion of infinitesimal). These
  systems cannot be controlled by digital controllers operating with a
  fixed sampling rate such as in \cite{CHR02}, since in this case their
  behaviors cannot be suitably discretized \cite{FR06,FPR08}.

\item Some well-known problems of --- possibly --- concurrent computation 
such as \emph{termination}, \emph{deadlocks}, and \emph{fairness} \cite{Fra86} 
can be considered as \emph{dual} problems to time advancement. In 
fact, they concern situations where some processes fail to advance 
their \emph{states}, while time keeps on flowing. Examples of these problems and their solutions are discussed with reference to a variety of formalisms introduced in Section \ref{sec:modernmodels}.
\end{itemize}

\noindent We can classify solutions to the time advancement problem into 
two categories: \emph{a priori} and \emph{a posteriori} methods.
In \emph{a priori} methods, the syntax or the semantics of the formal notation 
is restricted beforehand, in order to guarantee that the model 
of any system described with it will be exempt from time advancement 
problems. For instance, in some notations zero-time events are 
simply forbidden, or only finite sequences of them are allowed.

On the contrary, \emph{a posteriori} methods do not deal 
with time advancement issues until \emph{after} the system specification 
has been built; then, it is analyzed against a formal definition 
of time advancement in order to check that all of its actual 
behaviors do not incur into the time advancement problem. An \emph{a posteriori}
method may be particularly useful to spot possible 
risks in the behavior of the real system. For instance, in some 
cases oscillations exhibited by the mathematical model with a 
frequency that goes to infinity within a finite time interval, 
such as in the example above, may be the symptom of some instability 
in the modeled physical system, just in the same way as a physical 
quantity --- say, a temperature or a pressure --- that, in the mathematical 
model, tends to infinity within a finite time is the symptom 
of the risk of serious failure in the real system.

\subsection{Concurrency and Composition} \label{sec:concurrency}
Most real \emph{systems} --- as the term itself suggests --- are complex 
enough that it is useful, if not outright unavoidable, to model, 
analyze, and synthesize them as the composition of several subsystems. 
Such a composition/decomposition process may be iterated until 
each component is simple enough so that it can be analyzed in 
isolation.

Composition/decomposition, also referred to as \emph{modularization}, 
is one of the most general and powerful design principles in 
any field of engineering. In particular, in the case of --- sequential 
--- software design it produced a rich collection of techniques 
and language constructs, from subprograms to abstract data types.

The state of the art is definitely less mature when we come to 
the composition of concurrent activities. In fact, it is not 
surprising that very few programming languages 
deal explicitly with concurrency. It is well-known that the main 
issue with the modeling of concurrency is the \emph{synchronization} 
of activities (for which a plethora of more or less 
equivalent constructs are used in the literature: processes, 
tasks, threads, etc.) when they have to access shared resources 
or exchange messages.

The problem becomes even more intricate when the various activities 
are heterogeneous in nature. For instance, they may involve ``environment 
activities'' such as a plant or a vehicle to be controlled, and 
monitoring and control activities implemented through some hardware 
and software components. In such cases the time reference can 
be implicit for some activities, explicit for others; also, the 
system model might include parts in which time is represented 
simply as an ordering of events and parts that are described 
through a metric notion of time; finally, it might even be the 
case that different components are modeled by means of different 
time domains (discrete or continuous), thus producing hybrid 
systems.

Next, a basic classification of the approaches dealing with the 
composition of concurrent units is provided.

\subsubsection*{Synchronous vs.~Asynchronous Composition}
When composing concurrent modules there are two foremost ways 
of relating their temporal evolution: these are called synchronous 
and asynchronous composition.

Synchronous composition constraints state changes of the various 
units to occur at the very same time or at time instants that 
are strictly and rigidly related. Notice that synchronous composition 
is naturally paired with a discrete time domain, but meaningful 
exceptions may occur where the global system is synchronized 
over a continuous time domain.

Conversely, in an \emph{asynchronous} composition of parallel units, 
each activity can progress at a speed relatively unrelated with 
others; in principle there is no need to know in which state 
each unit is at every instant; in some cases this is even impossible: 
for instance, if we are dealing with a system that is geographically 
distributed over a wide area and the dynamics of some component 
evolves at a speed that is of the same order of magnitude as 
the light speed (more precisely, the state of a given component 
changes in a time that is shorter than the time needed to send 
information about the component's state to other components).

A similar situation occurs in totally different realms, such as the
world-wide stock market. There, the differences in local times between
locations all over the world make it impossible to define certain
states about the global market, such as when it is ``closed''.

For asynchronous systems, interaction between different components 
occurs only at a few ``meeting points'' according to suitably 
specified rules. For instance, the Ada programming language \cite{BB94} 
introduces the notion of \emph{rendez vous} between asynchronous 
tasks: a task owning a resource waits to grant it until it receives 
a request thereof; symmetrically a task that needs to access 
the resources raises a request (an \emph{entry call}) and waits 
until the owner is ready to accept it. When both conditions are 
verified (an entry call is issued and the owner is ready to accept 
it), the rendez vous occurs, that is, the two tasks are synchronized. 
At the end of the entry execution by the owner, the tasks split 
again and continue their asynchronous execution.

Many formalisms exist in the literature that aim at modeling 
some kind of asynchronous composition. Among these, Petri nets 
exhibit similarities with the above informal description of Ada's 
task system.

Not surprisingly, however, precisely formalizing the semantics 
of asynchronous composition is somewhat more complex than the 
synchronous one, and several approaches have been proposed in 
the literature. We examine some of them in Section \ref{sec:modernmodels}.

\subsection{Analysis and Verification Issues} \label{sec:analysis}
A fundamental feature of a formal model is its amenability to 
analysis; namely, we can probe the model of a system to be sure 
that it ensures certain desired features. In a widespread paradigm 
\cite{GJM02,Som04}, we call \emph{specification} the model under analysis, 
and \emph{requirements} the properties that the specification model 
must exhibit. The task of ensuring that a given specification 
satisfies a set of requirements is called \emph{verification}. Although 
this survey does not focus on verification aspects, we will occasionally 
deal with some related notions.

\subsubsection*{Expressiveness}
A fundamental criterion according to which formal languages can be
classified is their \emph{expressiveness}, that is, the possibility of
characterizing extensive classes of properties. Informally, a language
is more expressive than another one if it allows the designer to write
sentences that can more finely and accurately partition the set of
behaviors into those that satisfy or fail to satisfy the property
expressed by the sentence itself. Note that the expressiveness
relation between languages is a partial order, as there are pairs of
formal languages whose expressive power is incomparable: for each
language there exist properties that can be expressed only with the
other language. Conversely, there exist formalisms whose expressive
powers coincide; in such cases they are equivalent in that they can
express the very same properties. Expressiveness deals only with the
logical possibility of expressing properties; this feature is totally
different from other --- somewhat subjective, but nonetheless very
relevant --- characterizations such as conciseness, readability,
naturalness, and ease of use.

\subsubsection*{Decidability and Complexity}
Although in principle we might prefer the ``most expressive'' 
formalism, in order not to be restrained in what it be expressed, 
there is a fundamental trade-off between expressiveness and another 
important characteristic of a formal notation, namely, its \emph{decidability}. 
A certain property is \emph{decidable} for a formal language if 
there exists an algorithmic procedure that is capable of determining, 
for any specification written in that language, whether the property 
holds or not in the model. Therefore, the verification of decidable 
properties can be --- at least in principle --- a totally automated 
process. The trade-off between expressiveness and decidability 
arises because, when we increase the expressiveness of the language, 
we may lose decidability, and thus have to resort to semi-automated 
or manual methods for verification, or adopt \emph{partial} verification 
techniques such as testing and simulation. Here the term \emph{partial} refers to the fact that the analysis conducted with these techniques 
provides results that concern only a subset of all possible behaviors 
of the model under analysis.

While decidability is just a yes/no property, \emph{complexity} 
analysis provides, in the case when a given property is decidable, 
a measure of the computational effort that is required by an 
algorithm that decides whether the property holds or not for 
a model. The computational effort is typically measured in terms 
of the amount of memory or time required to perform the computation, 
as a function of the length of the input (that is, the size of 
the sentence that states it in the chosen formal language; see 
also Section \ref{sec:swview}).

\subsubsection*{Analysis and Verification Techniques}
There exist two large families of verification techniques: those 
based on \emph{exhaustive enumeration} procedures and those based 
on \emph{syntactic transformations} like deduction or rewriting, 
typically in the context of some axiomatic description. Although 
broad, these two classes do not cover, by any means, all the 
spectrum of verification algorithms, which comprises very different 
techniques and methods; here, however, we limit ourselves to 
sketching a minimal definition of these two basic techniques.

\emph{Exhaustive enumeration} techniques are mostly automated, and 
are based on exploration of graphs or other structures representing 
an operational model of the system, or the space of all possible 
interpretations for the sentence expressing the required property.

Techniques based on \emph{syntactic transformations} typically address 
the verification problem by means of logic deduction \cite{Men97}. 
Therefore, usually both the specification and the requirements 
are in descriptive form, and the verification consists of successive 
applications of some deduction schemes until the requirements 
are shown to be a logical consequence of the system specification.


\section{Historical Overview} \label{sec:historical}
In the rest of this article, in the light of the categories outlined 
in Section \ref{sec:dimensions}, we survey and compare a wide range of time models 
that have been used to describe computational aspects of systems.

This section presents an overview of the ``traditional'' 
models that first tackled the problem of time modeling, whereas 
Section \ref{sec:modernmodels} discusses some more ``modern'' formalisms. As stated 
in Section \ref{sec:metamodel}, we start from the description of \emph{formalisms}, 
but we will ultimately focus on their \emph{semantics} and, therefore, 
on what kind of \emph{temporal modeling} they allow.

Any model that aims at describing the ``dynamics'' of phenomena, 
or a ``computation'' will, in most cases, have some notion of 
time. The modeling languages that have been used from the outset 
to describe ``systems'', be they physical (e.g., moving objects, 
fluids, electric circuits), logical (e.g., algorithms), or even 
social or economic ones (e.g., administrations) are no exception 
and incorporate a more or less abstract idea of time.

This section presents the relevant features of the notion of 
time as traditionally used in three major areas of science and 
engineering: control theory (Section \ref{sec:dynamicalsys}),
electronics (Section \ref{sec:hwview}) and computer science (Section \ref{sec:swview}).
As the traditional modeling languages used in these disciplines have been widely studied 
and are well understood, we will only sketch their (well-known) 
main features; we will nonetheless pay particular attention to 
the defining characteristics of the notion of time employed in 
these languages, and highlight its salient points.

\subsection{Traditional Dynamical Systems} \label{sec:dynamicalsys}
A common way used to describe systems for control purposes in 
various engineering disciplines (mechanical, aerospace, chemical, 
electrical, etc.) is through the so-called state-space representation \cite{Kha95,SP05}.

The state-space representation is based on three key elements: 
a vector $\statex$ of \emph{state variables}, a vector $\inputu$ 
of \emph{input variables}, and a vector $\outputy$ of \emph{output variables}.
$\statex$, $\inputu$, and $\outputy$ are all \emph{explicit functions of time}, hence their 
values depend on the time at which they are evaluated and they 
are usually represented as $\statex(t)$, $\inputu(t)$, and $\outputy(t)$.\footnote{Another classic way of representing a dynamical system is through its \emph{transfer function}, which describes the input/output relationship of the system; unlike the state-space representation, the transfer function uses an \emph{implicit}, rather than explicit, notion of time. Despite its popularity and extensive use in the field of control theory, the transfer function has little interest in the modeling of computation, so we do not delve any further in its analysis.}

In the state-space representation the temporal domain is usually 
either continuous (e.g., $\reals$), or discrete (e.g., $\integers$). Depending 
on whether the temporal domain is $\reals$ or $\integers$, the relationship 
between $\statex$ and $\inputu$ is often expressed through differential 
or difference equations, respectively, e.g., in the following 
form: 
\begin{equation}\label{eq:star}
\begin{split}
   \dot{\statex}(t) & =   f(\statex(t), \inputu(t), t)   \\
   \statex(k+1) & = f(\statex(k), \inputu(k), k)
\end{split}
\end{equation}
where $t \in \reals$ and $k \in \integers$ (the relationship 
between $\outputy$ and the state and input variables is instead 
purely algebraic in the form $\outputy(t) = g(\statex(t), \inputu(t), t)$.

Given an initial condition $\statex(0)$, and fixed a function $\inputu(t)$, 
all functions $\statex(t)$ (or $\statex({k})$) if time is 
discrete) that are \emph{solutions} to the equations (\ref{eq:star}) represent 
the possible system behaviors. Notice that suitable constraints 
on the formalization of the system's dynamics are defined so that 
the derived behaviors satisfy some natural causality principles. 
For instance, the form of equations (\ref{eq:star}) must ensure that the 
state at time $t$ depends only on the initial state and on 
the value of the input in the interval $[0, t]$ (the future cannot modify the past).

Also, systems described through state-space equations are usually \emph{deterministic} 
(see Section \ref{sec:linear}), since the evolution of the state $\statex(t)$ 
is unique for a fixed input signal $\inputu(t)$ (and initial 
condition $\statex(0)$).\footnote{Notice that for a dynamical system described by equations such as (\ref{eq:star}) to be nondeterministic, the solution of the equation should be non-unique; this is usually ruled out by suitable hypotheses on the $f$ function \cite{Kha95}.} 
Therefore, dynamical system models typically assume a linear 
time model (see also the discussion in Section \ref{sec:linear}).

Moreover, time is typically treated quantitatively in these models, 
as the metric structure of the time domains $\reals$ or $\integers$ is 
exploited.

Notice also that often the first equation of (\ref{eq:star}) takes a simplified 
form:
\begin{equation*}
   \dot{\statex} \quad = \quad f(\statex, \inputu)
\end{equation*}
where the \emph{time} variable does not occur explicitly but is \emph{implicit} 
in the fact that $\statex$ and $\inputu$ are functions thereof. 
The time variable, of course, occurs explicitly in the solution 
of the equation. This is typical of time-invariant systems i.e., those 
systems that behave identically if the same ``experiment'' is 
translated along the time axis by a finite constant.

A typical example of continuous-time system is the electric circuit 
of Figure \ref{fig:RCcircuit}. A less common instance of discrete-time system is 
provided in the next example.

\begin{example}[Monodimensional Cellular Automata]
Let us consider a discrete-time family of dynamical systems called \emph{cellular automata},
where $\timedomain = \naturals$. More precisely, we consider the following instance,
named \emph{rule 110} by Wolfram \cite{Wol94}. The state domain is a bi-infinite string of binary values
$\stvar(k) = \ldots s_{i-2}(k) s_{i-1}(k) s_i(k) s_{i+1}(k) s_{i+2}(k) \ldots \in \{0,1\}^{2\omega}$,
and the output coincides with the whole state. The system is closed, since it has no input,
and its evolution is entirely determined by its initial state $s_i(0)$ ($i \in \integers$).

The dynamics is defined by the following equation, which determines 
the update of the state according to its value at the previous 
instant (starting from instant 1). 
\begin{equation*}
	 s_i(k+1) = 
	        \begin{cases}
				 1   &   \text{if } s_{i-1}(k) s_{i}(k) s_{i+1}(k) \in \{110, 101, 011, 010, 001\}  \\
				 0   &   \text{otherwise}
	        \end{cases}
\end{equation*}
Despite the simplicity of the update rule, the dynamics of such 
a system is highly complex; in particular it has been shown to 
be capable of universal computation \cite{Coo04}. 
\end{example}

Let us now discuss the main advantages in modeling an evolving 
process by means of a dynamical system. In doing so, perhaps 
we shift the point of view of system analysis from a ``control attitude''
towards a ``computing attitude''.

The foremost advantage of dynamical system models is probably 
that they borrow directly from the models used in physics. Therefore, 
they are capable of describing very general and diverse dynamic 
behaviors, with the powerful long-standing mathematical tools 
of differential (or difference) calculus. In particular, very 
different heterogeneous time models can be described within the 
same framework.\footnote{The recent literature of control theory also deals with hybrid systems where discrete and continuous time domains are integrated in the same system formalization \cite{vdSS00,Ant00,BBM98}.}
In a sense, many other formalisms for time modeling can be seen as a specialization of dynamical systems 
and can be reformulated in terms of state-space equations, including 
more computationally-oriented formalisms such as finite state 
automata and Turing machines.

The main limitations of the dynamical system models in describing 
timed systems lie in their being ``too detailed'' for some purposes. 
Being intrinsically operational and deterministic in most cases, 
such models provide complete descriptions of a system behavior, 
but are unsuitable for partial specifications or very high-level 
descriptions, which are instead a necessary feature in the early 
phases of development of a system (e.g., the requirements engineering 
phase in the development of software).

Moreover, since the time domain is usually a totally ordered 
metric set, dynamical systems are unsuitable for describing distributed 
systems where a notion of ``global time'' cannot be defined, and 
systems where the exact timing requirements are unspecified or 
unimportant. Some models that we will describe in the next sections 
try to overcome these limits by introducing suitable abstractions.

\subsection{The Hardware View} \label{sec:hwview}
One field in which the modeling of time has always been a crucial 
issue is (digital) electronic circuits design.

The key modeling issue that must be addressed in describing digital 
devices is the need to have \emph{different abstraction levels} 
in the description of the same system. More exactly, we typically 
have two ``views'' of a digital component. One is the \emph{micro 
view}, which is nearest to a physical description of the component. 
The other is the \emph{macro view}, where most lower-level details 
are abstracted away.

The \emph{micro view} is a low-level description of a digital component, 
where the basic physical quantities are modeled explicitly. System 
description usually partitions the relevant items into input, 
output, and state values. All of them represent physical quantities 
that vary continuously over time. Thus, the time domain is \emph{continuous}, 
and so is the state domain. More precisely, since we usually 
define an initialization condition, the temporal domain is usually \emph{bounded} 
on the left (i.e., $\reals_{\geq 0}$). Conversely, the state domain is often, but 
not always, restricted to a bounded subset $[L, U]$ of the whole 
domain $\reals$ (in many electronic circuits, for example, voltages 
vary from a lower bound of approximately 0V to an upper bound 
of approximately 5V).

Similarly to the case of time-invariant dynamical systems, time 
is generally \emph{implicit} in formalisms adopting the micro view. 
It is also \emph{metric} --- as it is always the case in describing 
directly physical quantities --- and fully \emph{asynchronous}, so 
that inputs may change at any instant of time, and outputs and 
states react to the changes in the inputs at any instant of time.

A simple operational formalism used to describe systems at the 
micro view is that of \emph{logic gates} \cite{KB04}, which can then 
be used to represent more complex digital components, with memory 
capabilities, such as \emph{flip-flops} and \emph{sequential machines}.

\begin{figure}[htb!]
	 \centering
	 \includegraphics{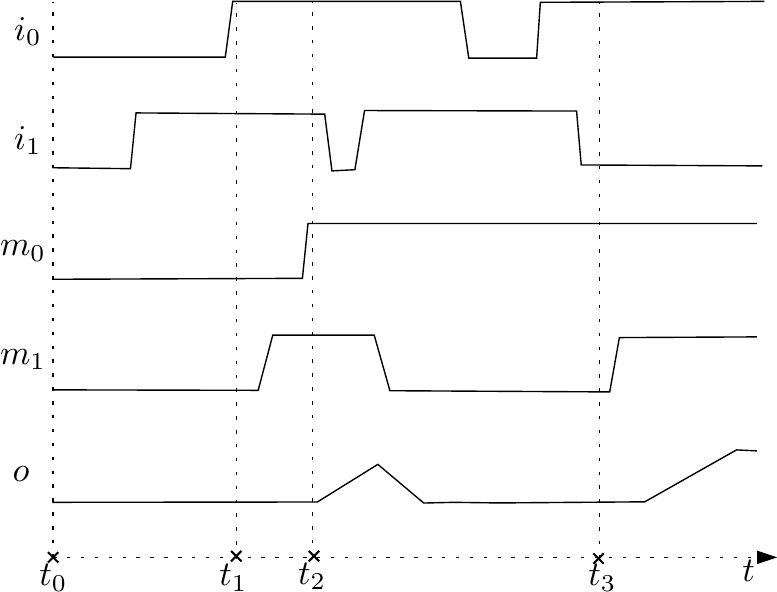}
	 \caption{A behavior of a sequential machine.}
	 \label{fig:sequential_machine_continuous}
\end{figure}

Figure \ref{fig:sequential_machine_continuous} shows an example of behavior of a sequential machine 
with two inputs $i_0$ and $i_1$, one output $o$, and two state values $m_0$ and $m_1$. 

The figure highlights the salient features of the modeling of 
time at the micro (physical) level: continuity of both time and 
state, and asynchrony (for example memory signals $m_0$ and $m_1$ can change 
their values at different time instants).

More precisely, Figure \ref{fig:sequential_machine_continuous} pictures a possible evolution of the 
state (i.e., the pair $\langle m_0, m_1 \rangle$) and of the output (i.e., signal $o$) of 
the sequential machine with respect to its input (i.e., the pair $\langle i_0, i_1 \rangle$).
For example, it shows that if all four signals $i_0, i_1, m_0, m_1$  are ``low''
(e.g., at time $t_0$), then the pair $\langle m_0, m_1 \rangle$ remains ``low''; 
however, if both input signals are ``high'' and the state is ``low''
(e.g., at time $t_1$), $m_1$ becomes ``high'' after a certain delay.
The output is also related to the state, in that $o$ is ``high'' when both
$m_0$ and $m_1$ are ``high'' (in fact, $o$ becomes ``high'' a little after $m_0$ and $m_1$ both become
``high'', as shown at time $t_3$ in the figure). Notice how the reaction to a change in the values 
of the input signals is not instantaneous, but takes a non-null 
amount of time (a \emph{propagation delay}), which depends on the 
propagation delays of the logic gates composing the sequential 
machine.

As the description above suggests, the micro view of digital circuits,
being close to the ``physical'' representation, is very detailed
(e.g., it takes into account the transient state that occurs after
variation in the inputs). However, if one is able to guarantee that
the circuit will eventually reach a stable state after a variation of
the inputs, and that the duration of the transient state is short with
respect to the rate with which input variations occur, it is
possible to abstract away the inner workings of the digital circuits,
and focus on the effects of a change in the inputs on the machine
state, instead.  In addition, it is common practice to represent the
``high'' and ``low'' values of signals in an abstract way, usually as
the binary values 1 (for ``high'') and 0 (for ``low'').  Then, we can
associate a sequential machine with a logic function that describes
the evolution of only the stable states. Table
\ref{tab:sequential_table} represents such a logic function where we
associate a letter to every possible stable configuration of the
inputs (column header) and the memory (row header), while the output
is instead simply defined to be 1 if and only if the memory has the
stable value 11. A blank cell in the table denotes an undefined
(``don't care'') behavior for the corresponding pair of current state
and current input. Then, the evolution in Figure
\ref{fig:sequential_machine_continuous} is compatible with the system
specification introduced by Table \ref{tab:sequential_table}.
\begin{table}[htb!]
  \begin{center}
	 \begin{tabular}{|l|cccc|}
	 \hline
                  &  $a(00)$   &  $b(01)$   &  $c(11)$   &  $d(10)$  \\
	 \hline
	 $A(00)$  &  $00$      &  $00$      &  $01$      &  $10$      \\
	 $B(01)$  &            &  $10$      &  $11$      &  $01$      \\
	 $C(11)$  &  $00$      &            &  $10$      &  $11$      \\
	 $D(10)$  &  $00$      &  $10$      &  $10$      &  $11$      \\
	 \hline
  \end{tabular}	 
  \end{center}
  \caption{A tabular description of the behavior of a sequential machine.}
  \label{tab:sequential_table}
\end{table}

Notice that by applying the above abstraction we discretized 
the state domain and assumed zero-time transitions. However, 
in the behavior of Figure \ref{fig:sequential_machine_continuous} the inputs $i_0$ and $i_1$ vary too quickly 
to guarantee that the component realizes the input/output function 
described by the table above. For example, when at instant $t_2$ both $m_0$
and $m_1$ become 1, memory signal $m_1$ does not have the time to become 
0 (as stated in Table \ref{tab:sequential_table}) before input $i_1$ changes anew. In addition, 
the output does not reach a stable state (and become 1) before state $m_1$
switches to 0. Thus, the abstraction of zero-time transition 
was not totally correct.

As the example suggests, full asynchrony in sequential machines 
poses several problems both at the modeling and at the implementation 
level. A very common way to avoid these problems, thus simplifying 
the design and implementation of digital components, is to synchronize 
the evolution of the components through a \emph{clock} (i.e., a physical signal that forces 
variations of other signals to occur only at its edges).

The benefits of the introduction of a clock are twofold: on the 
one hand, a clock rules out ``degenerate behaviors'', in which 
signal stability is never reached \cite{KB04}; on the other hand, 
it permits a higher-level view of the digital circuit, 
which we call the \emph{macro view}.

In the macro view, not only physical quantities are represented 
symbolically, as a combination of binary values; such values, 
in turn, are observed only when they have reached stable values. 
The time domain becomes discrete, too. In fact inputs are read 
only at periodic instants of time, while the state and outputs 
are simultaneously (and instantaneously) updated. Thus, their 
observation is \emph{synchronized} with a clock that beats the time 
periodically. Since we disregard any transient state, time is 
now a \emph{discrete} domain. In practice, we adopt the naturals $\naturals$
as time domain, whose origin matches the initialization instant 
of the system.

\begin{figure}[htb!]
	 \centering
	 \includegraphics{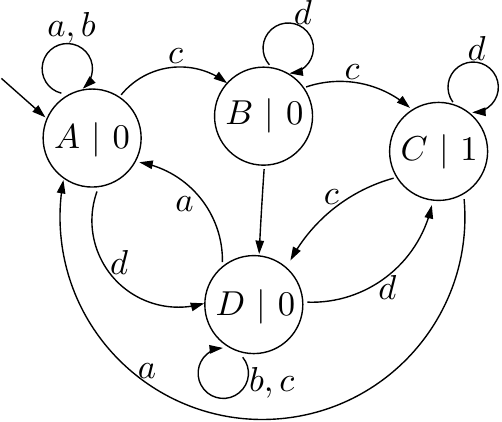}
	 \caption{A Moore machine.}
	 \label{fig:moore_machine}
\end{figure}

Typical languages that adopt ``macro'' time models are those that 
belong to the large family of abstract state machines \cite{Sip05,HMU00,MG87}. 
More precisely, the well-known Moore machines \cite{Moo56} and Mealy 
machines \cite{Mea55} have been used for decades to model the dynamics 
of digital components. For example, the Moore machine of Figure \ref{fig:moore_machine} 
represents the dynamics of the sequential machine implementing 
the logic function defined by Table \ref{tab:sequential_table}.
Every transition in the Moore machine corresponds to the 
elapsing of a clock interval; thus, the model abstracts away 
from all physical details, and focuses on the evolution of the 
system at clock ticks.

We will discuss abstract state machines and their notion of time 
in more detail in Section \ref{sec:synchronous}.

\subsection{The Software View} \label{sec:swview}
As mentioned above, abstract state machines such as the Moore 
machine of Figure \ref{fig:moore_machine} give a representation of digital components 
that is more ``computational-oriented'' and abstract than the 
``physics-oriented'' one of logic gates.

Traditionally, the software community has adopted a view of the 
evolution of programs over time that is yet more abstract.

In the most basic view of computation, time is not modeled at 
all. In fact, a software application implements a function of 
the inputs to the outputs. Therefore, the whole computational 
process is considered atomic, and time is absent from functional 
formalization. In other words, behaviors have no temporal characteristics 
in this basic view, but they simply represent input/output pairs 
for some computable function. An example of a formal language 
adopting such black-box view of computation is that of \emph{recursive functions},
at the roots of the \emph{theory of computation} \cite{Odi99,Rog87,BL74}.

A refinement of this very abstract way of modeling software would keep
track not only of the inputs and outputs of some computation, but also
of the whole sequence of discrete steps the computing device undergoes
during the computation (i.e., of the \emph{algorithm} realized by the
computation). More precisely, the actual time length of each step is
disregarded, assigning uniformly a unit length to each of them; this
corresponds to choosing the naturals $\naturals$ as time
domain. Therefore, time is discrete and bounded on the left:
the initial time 0 represents the time at which the computation
starts. The time measure represents the number of elementary
computational steps that are performed through the computation. Notice
that no form of concurrency is allowed in these computational
models, which are strictly \emph{sequential}, that is each step is
followed uniquely by its successor (if any).

\begin{figure}[htb!]
	 \centering
	 \includegraphics[scale=1.2]{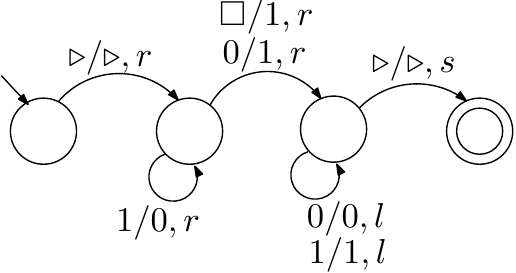}
	 \caption{A Turing machine computing the successor function. $\triangleright$ denotes the origin of the Turing machine tape, $\Box$ denotes the blank symbol; a double circle marks a halting state; in every transition, $I/O,M$ denotes the symbol read on the tape upon taking the transition ($I$), the symbol ($O$) written on the tape in place of $I$, and the way ($M$) in which the tape head is moved ($l$ for ``left'', $s$ for ``stay'', and $r$ for ``right'').}
	 \label{fig:turing_machine}
\end{figure}

Turing Machines \cite{Pap94,Sip05,HMU00,MG87} are a classic formalism 
to describe computations (i.e., algorithms). For example, the 
Turing machine of Figure \ref{fig:turing_machine} describes an algorithm to compute 
the successor function for a binary input (stored with the least 
significant bit on the left, that is in a ``little-endian'' fashion).

For a given Turing machine $M$ computing a function $f$, or 
any other abstract machine for which it is assumed that an elementary 
transition takes a time unit to execute, by associating the number 
of steps from time 0 until the reaching of a halting state --- if 
ever --- we may define a \emph{time complexity} function $T_{M}(n)$, whose argument
$n$ represents the \emph{size} of the data input to $f$, and whose value is the maximum number of steps required 
to complete the computation of $f$ when input data has size $n$.\footnote{As a particular case, if $M$'s computation never reaches a halting state we conventionally define $T_{M}(n) = \infty$.}
For example, the computational complexity $T_{\mathrm{succ}}(n)$ of the Turing machine of Figure \ref{fig:turing_machine} is proportional to the length $n$ of the input string.

In the software view the functional behavior of a computation 
is normally separated from its timed behavior. Indeed, while 
the functional behavior is studied without taking time into account, 
the modeling of the timed behavior focuses only on the number 
of steps required to complete the computation. In other words, 
functional correctness and time complexity analysis are usually 
separated and adopt different techniques.

In some sense the software view of time models constitutes a 
further abstraction of the macro hardware view. In particular, 
the adoption of a discrete time domain reflects the fact that 
the hardware is what actually performs the computations formalized 
by means of a software model. Therefore, all the abstract automata 
that are used for the macro modeling of hardware devices are 
also commonly considered models of computation.


\section{Temporal Models in Modern Theory and Practice} \label{sec:modernmodels}
The growing importance and pervasiveness of computer systems 
has required the introduction of new, richer, and more expressive 
temporal models, fostering their evolution from the basic ``historical'' 
models of the previous section. This evolution has inevitably 
modified the boundaries between the traditional ways of modeling 
time, often making them fuzzy. In particular, this happened with 
heterogeneous systems, which require the combination of different abstractions 
within the same model.

This section shows how the aforementioned models have been refined 
and adapted in order to meet more specific and advanced specification 
needs. These are particularly prominent in some classes of systems, 
such as hybrid, critical, and real-time systems \cite{HM96}. As we 
discussed above in Section \ref{sec:introduction}, these categories are independent 
but with large areas of overlap.

\begin{table}[!ht]
\centering
\begin{scriptsize}
\begin{tabular}{|p{5cm}|c|c|}
\hline 
\textbf{Keywords} & \textbf{Dimension} & \textbf{Section} \\
\hline 
discrete, dense, continuous, granularity & \emph{Discrete vs.~Dense} & \ref{sec:discrete} \\
\hline 
qualitative, quantitative, metric(s) & \emph{Ordering vs.~Metric} & \ref{sec:ordering} \\
\hline 
linear, branching, (non)deterministic & \emph{Linear vs.~Branching} & \ref{sec:linear} \\
\hline 
implicit(ly), explicit(ly) & \emph{Implicit vs.~Explicit} & \ref{sec:implicit} \\
\hline 
(non)-Zeno, fairness, deadlock(ed) & \emph{Time Advancement} & \ref{sec:timeadvancement} \\
\hline 
composing, composition, concurrency, synchrony, synchronous(ly), asynchronous(ly) & \emph{Concurrency and Composition} & \ref{sec:concurrency} \\
\hline 
analysis, tool(set), verification, decision procedure & \emph{Analysis and Verification} & \ref{sec:analysis} \\
\hline 
\end{tabular}
\end{scriptsize}
\caption{Keyword references to the ``Dimensions'' of Section \ref{sec:dimensions}.}
\label{tab:keywords}
\end{table}

As in the historical overview of Section \ref{sec:historical}, the
main features of the models presented in this section are discussed
along the dimensions introduced in Section \ref{sec:dimensions}. Such
dimensions, however, have different relevance for different
formalisms; in some cases a dimension may even be unrelated with some
formalism. For this reason we avoid a presentation in the style of a
systematic ``tabular'' cross-reference $<\text{Formalism}/\text{dimension}>$; rather,
to help the reader match the features of a formalism with the
coordinates of Section \ref{sec:dimensions}, we highlight the portions
of the text where a certain dimension is specifically discussed by
graphically emphasizing (in small caps) some related
keywords. The correspondence between keywords and dimensions is shown
in Table \ref{tab:keywords}. Also, for the sake of conciseness, we do
not repeat features of a \emph{derived} formalism that are inherited
unaffected from the ``parent'' notation.

\subsubsection*{The Computer- and System-Centric Views}
As a preliminary remark we further generalize the need of adopting 
and combining different views of the same system and of its heterogeneous 
components. Going further --- and, in some sense, back --- in the path 
described in Section \ref{sec:historical}, which moved from the micro to the macro 
view of hardware, and then to the software view, we now distinguish 
between a \emph{computer-centric} and \emph{a system-centric} view. 
As the terms themselves suggest, in a computer-centric view attention 
is focused on the computing device and its behavior, which may 
involve interaction with its environment through I/O operations; 
in a system-centric view, instead, attention is on a whole collection 
of heterogeneous components, and computing devices --- hardware 
and software --- are just a subset thereof. Most often, in such 
systems the dynamics of the components range over widely different 
time scales and time granularities (in particular continuous 
and discrete components are integrated).

In a \emph{computer-centric} view we consider systems where 
time is inherently discrete, and which can be described with 
a (finite-)state model. Moreover, we usually adopt a strictly 
synchronous model of concurrency, where the global synchrony 
of the system is given by the global clock ticking. Nondeterminism 
is also often adopted to model concurrent computations at an 
abstract level.

Another typical feature of this view is the focus on the ease 
of --- possibly automated --- analysis to validate some properties; 
in general, it is possible and preferred to restrict and abstract 
away from many details of the time behavior in favor of a decidable 
formal description, amenable to automated verification.

An example of computer-centric view is the design and analysis 
of a field bus for process control: the attention is focused 
on discrete signals coming from several sensors and on their 
proper synchronization; the environment that generates the signals 
is ``hidden'' by the interface provided by the sensors.

Conversely, in the \emph{system-centric} view, the aim is to model, 
design, and analyze the whole system; this includes the process 
to be controlled, the sensors and actuators, the network connecting 
the various elements, the computing devices, etc.

In the \emph{system-centric} view, according to what kind of application 
domain we consider, time is sometimes continuous, and sometimes 
discrete. The concurrency model is often asynchronous, and the 
evolution of components is usually deterministic. For instance, 
a controlled chemical process would be described in terms of 
continuous time and asynchronous deterministic processes; on 
the other hand a logistic process --- such as the description 
of a complex storage system --- would be probably better described 
in terms of discrete time. Finally, the system-centric view puts 
particular emphasis on input/output variables, modular divisions 
among components, and the resulting ``information flow'', similarly 
to some aspects of dynamical systems. Thus, the traditional division 
between hardware and software is blurred, in favor of the more 
systemic aspects.

In practice, no model is usually taken to be totally computer-centric 
or system-centric; more often, some aspects of both views are 
united within the same model, tailored for some specific needs.

\paragraph{}
The remainder of this section presents some broad classes of 
formal languages, in order to discuss what kind of temporal models 
they introduce, and what kind of systems they are suitable to 
describe.

We first analyze a selected sample of operational formalisms.  Then,
we discuss descriptive formalisms based on logic, and devote
particular attention to some important ones. Finally, we present
another kind of descriptive notations, the algebraic formalisms, that
are mostly timed versions of successful untimed formal languages and
methods.

To discuss some features of the formalisms surveyed we will 
adopt a simple running example based on a resource allocator. 
Let us warn the reader, however, that the various formalizations 
proposed for the running example do not aim at being different 
specifications of the same system; on the contrary, the semantics 
may change from case to case, according to which features of 
the formalism we aim at showing in that particular instance.

\subsection{Operational Formalisms} \label{sec:operational}
We consider three broad classes of operational formalisms: synchronous 
state machines, Petri nets as the most significant exponent of 
asynchronous machines, and heterogeneous models.

\subsubsection{Synchronous Abstract Machines} \label{sec:synchronous}
In Section \ref{sec:historical} we presented some classes of (finite-)state 
machines that have a synchronous behavior. As we noticed there, 
those models are mainly derived from the synchronous ``macro'' 
view of hardware digital components, and they are suitable to 
describe ``traditional'' \emph{sequential} computations.

The natural evolution of those models, in the direction of increasing
complexity and sophistication, considers \emph{concurrent} and
\emph{reactive} systems. These are, respectively, systems where
different components operate in parallel, and open systems whose
ongoing interaction with the environment is the main focus, rather
than a static input/output relation. The models presented in this
section especially tackle these new modeling needs.

\paragraph{Infinite-Word Finite-State Automata.}
Perhaps the simplest extension of automata-based formalisms to 
deal with reactive computations consists in describing a semantics 
of these machines over infinite (in particular, denumerable) 
sequences of input/output symbols. This gives rise to finite-state 
models that are usually called ``automata on infinite words'' (or $\omega$-words).
The various flavors of these automata differ in how they define 
acceptance conditions (that is, how they distinguish between 
the ``good'' and ``bad'' interactions with the environment) and 
what kind of semantic models they adopt.

Normally these models are defined in a nondeterministic version, 
whose transition relation $\delta \subseteq \Sigma \times \sdomain \times \sdomain$
(where $\Sigma$ is the input alphabet, and $\sdomain$ is the state space) associates
input symbol, current state and next state. 
Thus, for the same pair $\langle \sigma, s \rangle$ of input symbol and current state,
more than one next state $n$ may be in relation with it, that is, the automaton can ``choose'' 
any of the next states in the set $\{n \:|\: \langle \sigma, s, n \rangle \in \delta \}$.
Nondeterminism and infinite words require the definition of different, 
more complex acceptance conditions than in the deterministic, 
finite word case. For instance, the B\"uchi acceptance condition 
is defined through a set of final states, some of which must 
be visited infinitely often in at least one of the nondeterministically-chosen 
runs \cite{Var96}. Other acceptance conditions are defined, for instance, 
in Rabin automata, Streett automata, parity automata, Muller 
automata, tree automata, etc. \cite{Tho90}.

As an example of use of infinite-word automata, let us model 
a simple resource manager. Before presenting the example, however, 
we warn the reader that we are not interested in making the resource 
manager as realistic as possible; rather, as our aim is to show 
through small-sized models the most relevant features of the 
formalisms presented, for the sake of brevity we introduce simplifications 
that a real-world manager would most probably avoid.

The behavior of the resource manager is the following: Users 
can issue a request for a resource either with high priority 
($\hpr$) or with low priority ($\lpr$). Whenever the resource 
is free and a high-priority request is raised, the resource is 
immediately granted and it becomes occupied. If it is free and 
a low-priority request is received, the resource is granted after 
two time units. Finally, if a high-priority request is received 
while the resource is granted, it will be served as soon as the 
resource is released, while a low-priority request will be served 
two instants after the resource is released. Further requests 
received while the resource is occupied are ignored.

\begin{figure}[htb!]
	 \centering
	 \includegraphics{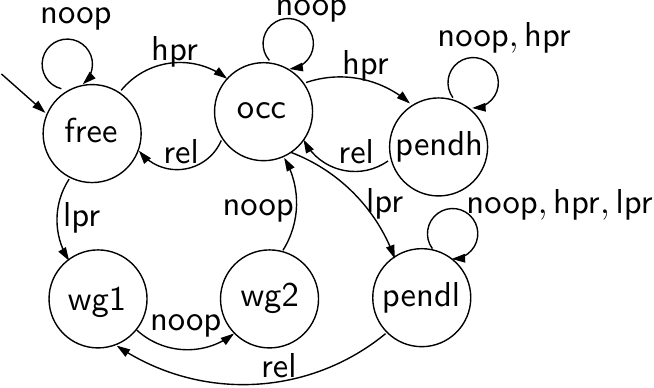}
	 \caption{A resource manager modeled by an infinite-word finite-state automaton.}
	 \label{fig:allocator}
\end{figure}

The above behavior can be modeled by the automaton of Figure \ref{fig:allocator}, 
where the various requests and grant actions define the input 
alphabet (and $\noop$ defines a ``waiting'' transition); note 
that the automaton is actually deterministic. We assume that 
all states are accepting states.

Let us analyze the infinite-word finite-state automaton models 
with respect to our coordinates. First of all, these models can 
be considered as mainly ``computer-centric'', focusing on simplicity 
and abstractness. In particular, from the point of view of the 
computer scientist, they are particularly appealing, as they 
allow one to reason about time in a highly simplified way.

There is no explicit notion of quantitative time. As usual, however, 
a simple \kw{metric} is implicitly defined by associating a time unit 
with the execution of a single transition; thus time is inherently \kw{discrete}. 
For example, in Figure \ref{fig:allocator}, we measure \kw{implicitly} the two time units after 
which a low priority request is granted, by forcing the path 
from the request $\lpr$ to $\occ$ to pass through two intermediate 
states via two ``wait'' transitions $\noop$.

The simplicity of the time model makes it amenable to automated 
\kw{verification}. Various techniques have been developed to analyze 
and verify automata, the most successful of whom is probably \emph{model checking} \cite{CGP00}.
(See also Section \ref{sec:duallanguage}).

The \kw{nondeterministic} versions of these automata are particularly 
effective for characterizing multiple computation paths. In defining 
its formal semantics one may exploit a \kw{branching} time model. 
There are, however, relevant examples of nondeterministic automata 
that adopt a \kw{linear} time model, B\"uchi automata being the most 
noticeable instance thereof. In fact, modeling using linear time 
is usually considered more intuitive for the user; for instance, 
considering the resource manager described above, the linear 
runs of the automaton naturally represent the possible sequences 
of events that take place in the manager. This intuitiveness 
was often considered to be traded off with amenability to automatic 
verification, since the first model checking procedures were 
more efficient with branching logic \cite{CGP00}. Later progresses 
have shown, however, that this trade off is often fictitious, 
and linear time models may be endowed with efficient verification 
procedures \cite{Var01}.

When \kw{composing} multiple automata in a global system we must face 
the problem of \kw{concurrency}. The two most common concurrency models 
used with finite automata are \emph{synchronous} concurrency and \emph{interleaving} 
concurrency. 

\begin{itemize}
\item In \emph{synchronous concurrency}, concurrent transitions of different 
composed automata occur simultaneously, that is the automata 
evolve with the same ``global'' time. This approach is probably 
the simpler one, since it presents a global, unique vision of 
time, and is more akin to the ``synchronous nature'' of finite-state 
automata. Synchronous concurrency is pursued in several languages 
that constitute extensions and specializations of the basic infinite-word 
finite-state automaton, such as Esterel \cite{BG92} and Statecharts (see below).

\item In \emph{interleaving concurrency}, concurrent transitions 
are ordered arbitrarily. Then any two global orderings of the 
transitions that differ only for the ordering of concurrent transitions 
are considered equivalent. Interleaving semantics may be regarded 
as a way to introduce a weak notion of concurrency in a strictly 
synchronous system. The fact that interleaving introduces partially 
ordered transitions weakens however the intuitive notion of time 
as a total order. Also, the natural correspondence between the 
execution of a single transition and the elapsing of a time unit 
is lost and \emph{ad hoc} rules are required to restate a time metric 
based on the transition execution sequence.

Another problem introduced by adopting an interleaving semantic 
model lies in the \kw{fairness} requirement, which prescribes that 
every concurrent request eventually gets satisfied. Usually, 
fairness is enforced explicitly \emph{a priori} in the 
composition semantic. 
\end{itemize}

The main strength of the infinite-word finite-state automata 
models, i.e., their simplicity, constitutes also their main limitation. 
When describing physical systems, adopting a strictly synchronous 
and discrete view of time might be an obstacle to a ``natural'' 
modeling of continuous processes, since discretization may be 
too strong of an abstraction. In particular, some properties 
may not hold after discretization, such as periodicity if the 
duration of the period is some irrational constant, incommensurable 
with the duration of the step assumed in the discretization. 

Moreover, it is very inconvenient to represent heterogeneous systems
with this formalism when different components run at highly different
speeds and the time \kw{granularity} problem arises. In more technical
terms, for this type of models it is rather difficult to achieve
\emph{compositionality} \cite{AFH96,AH92}.

\paragraph{Statecharts.}
Statecharts are an automata-based formalism, invented by \linebreak
David~Harel~\cite{Har87}. They are a quite popular tool in the
software engineering community, and a version thereof is part of the
UML standard \cite{UML05,UML04}.

In a nutshell, Statecharts are an enrichment of classical finite-state 
automata that introduces some mechanisms for hierarchical abstraction 
and parallel composition (including synchronization and communication 
mechanisms). They may be regarded as an attempt to overcome some 
of the limitations of the bare finite-state automaton model, 
while retaining its advantages in terms of simplicity and ease 
of graphical representation. They assume a synchronous view of 
communication between parallel processes.

\begin{figure}[htb!]
	 \centering
	 \includegraphics[scale=1.15]{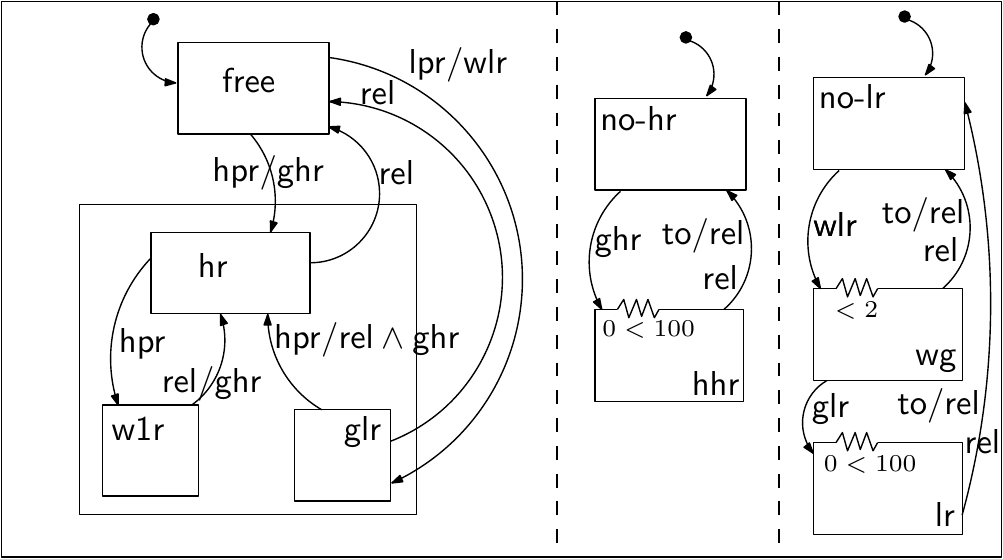}
	 \caption{A resource manager modeled through a Statechart.}
	 \label{fig:statechart}
\end{figure}

Let us use the resource manager running example to illustrate 
some of Statecharts' features; to this purpose we introduce some 
modifications to the initial definition. First, after any request has 
been granted, the resource must be released within 100 time units. 
To model such \kw{metric} temporal constraints we associate a \emph{timeout} 
to some states, namely those represented with a short squiggle 
on the boundary (such as $\hhr$ or $\wg$ in Figure \ref{fig:statechart}).

Thus, for instance, the transition that exits state $\hhr$ must be
taken within 100 time units after $\hhr$ has been entered: if no
$\rel$ event has been generated within 100 time units, the timeout
event $\too$ is ``spontaneously-generated'' exactly after 100 time
units.\footnote{Note that there are in fact two transitions from state
  $\hhr$ to state $\nohr$, one that is labeled $\too/\rel$, and one
  that is labeled $\rel$; they are represented in Figure
  \ref{fig:statechart} with a single arc instead of two separate ones
  for the sake of readability. The transition labeled $\too/\rel$
  indicates that when the timeout expires (the $\too$ event), a $\rel$
  event is triggered, which is then sensed by the other parts of the
  Statechart, hence producing other state changes (for example from
  $\glr$ to $\free$).}  Conversely the lower bound of 0 in the same
state indicates that the same transition cannot be taken
immediately. We use the same mechanism to model the maximum amount of
time a low-priority request may have to wait for the resource to
become available; in this case, with respect to the previous example,
we allow the low-priority request to be granted immediately,
nondeterministically. Notice that modeling time constraints using
timeouts (and exit events) implies an \kw{implicit} modeling of a
global system time, with respect to which timeouts are computed, just
like in finite-state automata.  In fact, timeouts can be regarded as
an enrichment of the discrete finite state automaton model with a
\kw{continuous} feature.

The example of Figure \ref{fig:statechart} exploits Statecharts'
so-called ``AND (parallel) composition'' to represent three logically
separable components of the system, divided by dashed lines. The
semantics of AND composition is obtained as the Cartesian product
construction,\footnote{In fact, the semantics of the AND composition
  of submachines in Statecharts differs slightly from the classic
  notion of Cartesian product of finite-state machines; however, in
  this article we will not delve any further in such details, and
  instead refer the interested reader to \cite{Har87} for a deeper
  discussion of this issue.} and it is usually called
\kw{synchronous composition};\footnote{We warn the reader that the
  terminology often varies greatly among different areas; for instance
  \cite{CL99} names the Cartesian product composition ``completely
  asynchronous''.} however, Statecharts' graphical representation
avoids the need to display all the states of the product construction,
ameliorating the readability of a complex specification. In particular
in our example, we choose to allow one pending high-priority request
to be ``enqueued'' while the resource is occupied; thus the leftmost
component is a finite-state automaton modeling whether the resource is
free, serving a high-priority request with no other pending requests
(state $\hr$), or with one pending request (state $\woner$), or
serving a low-priority request (state $\glr$).

Since in Statecharts all transition events --- both input and output 
--- are ``broadcast'' over the whole system, labeling different 
transitions with the same name enforces synchronization between 
them. For instance, whenever the automaton is in the global state $\langle \woner, \hhr, \nolr \rangle$, 
a release event $\rel$ triggers the global state to become $\langle \hr, \nohr, \nolr \rangle$, 
and then cascading immediately to $\langle \hr, \hhr, \nolr \rangle$, 
because of the output event $\ghr$ triggered by the transition 
from $\woner$ to $\hr$. Note that we are implicitly assuming, 
in the example above, that $\ghr$ and $\wlr$ are ``internal events'',
i.e., they do not occur spontaneously in the environment 
but can only be generated internally for synchronization.

\begin{figure}[htb!]
	 \centering
	 \includegraphics{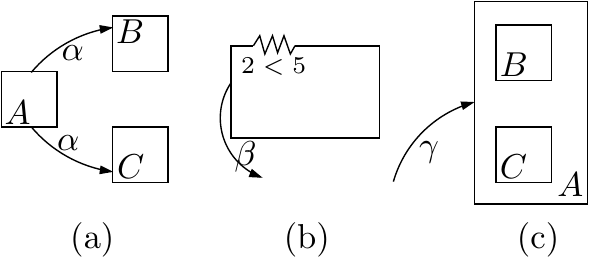}
	 \caption{Nondeterminism in Statecharts.}
	 \label{fig:statecharts-nondet}
\end{figure}

\kw{Nondeterminism} can arise in three basic features of Statecharts 
models. First, we have the ``usual'' nondeterminism of two mutually 
exclusive transitions with the same input label (such as in Figure \ref{fig:statecharts-nondet}(a)). 
Second, states with timeout are exited nondeterministically \emph{within} 
the prescribed bounds (Figure \ref{fig:statecharts-nondet}(b)). Third, Statechart modules 
may be composed with ``XOR composition'', that represents a nondeterministic 
choice between different modules (Figure \ref{fig:statecharts-nondet}(c)). 

The popularity of Statecharts has produced an array of different 
\kw{analysis tools}, mostly automated. For instance \cite{HLNPPSST90,BDW00,GTBF03}.

While overcoming some of the limitations of the basic finite-state 
automata models, Statecharts' rich syntax often hides subtle semantic 
problems that instead should be better exposed to avoid inconsistencies 
and faults in specifications. In fact, over the years several 
researches have tried to define formally the most crucial aspects 
of the temporal semantics of Statecharts. The fact itself that 
different problems were unveiled only incrementally by different 
contributors is an indication of the difficulty of finding a 
comprehensive, intuitive, non-ambiguous semantics to an apparently 
simple and plain language. We discuss here just a few examples, 
referring the interested reader to \cite{HPSS87,PS91,vdB94,HN96} 
for more details.

The apparently safe ``perfect \kw{synchrony}'' assumption --- the assumption 
that all transition events occur simultaneously --- and the global 
``broadcast'' availability of all events --- which are therefore 
non local --- generate some subtle difficulties in obtaining a 
consistent semantics. Consider for instance the example of Figure \ref{fig:statechart}, 
and assume the system is in the global state $\langle \glr, \nohr, \lr \rangle$. 
If a high-priority request takes place, and thus a $\hpr$ event 
is generated, the system shifts to the state $\langle \hr, \nohr, \lr \rangle$ in zero time.
Simultaneously, the taken transition triggers the events $\rel$ and $\ghr$.
If we allow a zero-time residence in states, the former event moves the system
to $\langle \hr, \nohr, \nolr \rangle$, representing the low-priority request being forced to release 
the resource. Still simultaneously, the latter $\ghr$ event 
triggers the transition from $\nohr$ to $\hhr$ in the middle sub-automaton.
This is in conformity with our intuitive requirements; however the same $\rel$ generated event also triggers 
the first sub-automaton to the state $\free$, which is instead 
against the intuition that suggests that the event is only a 
message sent to the other parts of the automaton.

If we refine the analysis, we discover that the picture is even 
more complicated. The middle automaton is in fact in the state $\hhr$, 
while the time has not advanced; thus we still have the $\rel$ 
event available, which should immediately switch the middle automaton 
back to the state $\nohr$. Besides being intuitively not 
acceptable, this is also in conflict with the lower bound on 
the residence time in $\hhr$. Moreover, in general we may end 
up having multiple XOR states occupied at the same time. Finally, 
it is not difficult to conceive scenarios in which the simultaneous 
occurrence of some transitions causes an infinite sequence of 
states to be traversed, thus causing a \kw{Zeno} behavior.

How to properly disentangle such scenarios is not obvious. A 
partial solution would be, for instance, to avoid instantaneous 
transitions altogether, attaching a non-zero time to transitions 
and forcing an ordering between them or, symmetrically, to disallow 
a zero-time residence in states. This (partially) asynchronous 
approach is pursued for instance in Timed Statecharts \cite{KP92}, 
or in other works \cite{Per93}. Alternatively, other solutions disallow 
loops of zero-time transitions, but accept a finite number of 
them (for instance, by ``consuming'' each event spent by a transition \cite{HN96});
the Esterel language, which is a ``relative'' of Statecharts', follows this approach.

\paragraph{Timed and Hybrid Automata.}
As we discussed above, the strictly discrete and synchronous view of
finite-state automata may be unsuitable to model adequately and
compositionally processes that evolve over a dense domain.
Statecharts try to overcome these problems by adding some continuous
features, namely timeout states. Timed and hybrid automata push this
idea further, constituting models, still based on finite-state
automata, that can manage continuous variables. Let us first discuss
timed automata.

\emph{Timed automata} enrich the basic finite-state automata with 
real-valued \emph{clock} variables. Although the name ``timed automata'' 
could be used generically to denote automata formalisms where 
a description of time has been added (e.g., \cite{LV96,AH96,Arc00}), 
here we specifically refer to the model first proposed by Alur 
and Dill \cite{AD94}, and to its subsequent enrichments and variations. 
We refer the reader to Alur and Dill's original paper \cite{AD94} and to \cite{BY04} 
for a formal, detailed presentation.

In timed automata, the total state is composed of two parts: 
a finite component (corresponding to the state of a finite automaton, 
which is often called \emph{location}), and a continuous one represented 
by a finite number of positive real values assigned to variables 
called \emph{clocks}. The resulting system has therefore an \emph{infinite} 
state space, since the clock components take value in the infinite 
set $\reals_{\geq 0}$. The evolution of the system is made of alternating phases 
of instantaneous synchronous discrete ``jumps'' and continuous 
clock increases. More precisely, whenever a timed automaton sits 
in some discrete state, each clock variable $x$ increases as 
time elapses, that is it evolves according to the dynamic equation $\dot{x} = 1$, 
thus effectively measuring time. External input events cause 
the discrete state to switch; during the transition some clock 
variables may be reset to zero instantaneously. Moreover, both 
discrete states and transitions may have attached some constraints 
on clocks; each constraint must be satisfied while sitting in 
the discrete state, and when taking the transition, respectively.\footnote{The original Alur and Dill's formalization \cite{AD94} permitted constraints only on transitions; however, adding constraints to locations as well is a standard extension that does not impact on the salient features of the model (expressiveness, in particular) \cite{BY04}.}

\begin{figure}[htb!]
	 \centering
	 \includegraphics{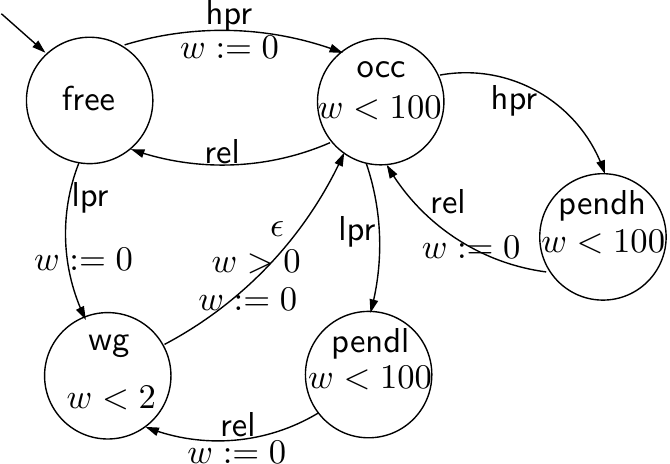}
	 \caption{A resource manager modeled through a timed automaton.}
	 \label{fig:timed_automaton}
\end{figure}

To illustrate this notation, let us model the resource manager 
example through a timed automaton. We modify the system behavior 
of the Statechart example, by disallowing high-priority requests 
to preempt low-priority ones; moreover, let us assume that one 
low-priority request can be ``enqueued'' waiting for the resource 
to become free. The resulting timed automaton --- using a single 
clock $w$ --- is pictured in Figure \ref{fig:timed_automaton}.

The semantics of a timed automaton is usually formally defined 
by means of a timed transition system. The ``natural'' semantics 
is the \emph{timed} semantics, which exactly defines the possible 
runs of one automaton over sequences of input symbols. More precisely, 
each symbol in the input sequence is paired with a \emph{timestamp} 
that indicates the absolute time at which the symbol is received. 
Then, a run is defined by a sequence of total states (each one 
a pair $\langle \text{location, clock value} \rangle$ of 
the automaton, which evolve according to the timestamped input 
symbols, in such a way that, for every pair of consecutive states
$\langle l_i, c_i \rangle \xrightarrow{\mathsf{in}, ts} \langle l_{i+1}, c_{i+1} \rangle$
in the run the constraints on the locations and the transition 
are met. For instance, the automaton of Figure \ref{fig:timed_automaton} may go through 
the following run:
\begin{equation*}
 \langle \mathsf{free}, 0 \rangle \xrightarrow{\mathsf{hpr}, 4.7} \langle \mathsf{occ}, 0 \rangle \xrightarrow{\mathsf{lpr}, 53.9} \langle \mathsf{pendl}, 49.2 \rangle \xrightarrow{\mathsf{rel}, 64} \langle \mathsf{wg}, 0 \rangle \xrightarrow{\epsilon, 65.1} \langle \mathsf{occ}, 0 \rangle \cdots
\end{equation*}
In the run above state location $\occ$ is entered at time 4.7 
and, since the corresponding transition resets clock $w$, the 
new state becomes $\langle \occ, 0 \rangle$; then, at time 
53.9 (when clock $w$ has reached value 49.2), location $\occ$
is exited and $\mathsf{pendl}$ is entered (this time, the clock $w$ 
is not reset), which satisfies the constraint $w < 100$
of location $\occ$, and so on.

Timed semantics introduces a \kw{metric} treatment of time through 
timestamps. Notice that, in some sense, the use of timestamps 
introduces ``two different notions of time'': the inherently \kw{discrete} 
one, given by the position $i$ in the run/input sequence, which 
defines a total ordering on events, and the \kw{continuous} and metric 
one, recorded by the timestamps and controlled through the clocks. 
This approach, though simple in principle, somewhat sacrifices 
naturalness, since a complete time modeling is no more represented 
as a unique flow but is two-fold.

Other, different semantics of timed automata have been introduced 
and analyzed in the literature. Subtle differences often arise 
depending on which semantics is adopted; for instance, interval-based 
semantics interprets timed automata over piecewise-constant functions 
of time, and the change of location is triggered by discontinuities 
in the input \cite{AFH96,ACM02,Asa04}.

Let us consider a few more features of time modeling for timed 
automata. 
\begin{itemize}
\item While timed automata are in general \kw{nondeterministic}, their 
semantics is usually defined through \kw{linear} time models, such 
as the one outlined above based on run sequences. Moreover, deterministic 
timed automata are strictly less expressive than nondeterministic 
ones, but also more amenable to automated verification, so they 
may be preferred in some practical cases.

\item \emph{Absolute time is} \kw{implicitly} assumed in the model 
and becomes apparent in the timestamps associated with the input 
symbols. The \emph{relative time} measured by clocks, however, is \kw{explicitly} 
measured and set.

\item Timed automata may exhibit \kw{Zeno} \emph{behaviors}, when distances 
between times at which transitions in a sequence are taken become 
increasingly smaller, accumulating to zero. For instance, in 
the example of Figure \ref{fig:timed_automaton}, the two transitions $\hpr$ and $\rel$ 
may be taken at times $1, 1+2^{-1}, 1+2^{-1}+2^{-2}, \ldots, \Sigma_{k = 0}^{n} 2^{-k}, \ldots$, so that the absolute time would accumulate at $\Sigma_{k = 0}^{\infty} 2^{-k} = 2$. Usually, these Zeno behaviors are ruled out \emph{a priori} 
in defining the semantics of timed automata, by requiring that 
timestamped sequences are acceptable only when the timestamp 
values are unbounded.

Moreover, in Alur and Dill's formulation \cite{AD94} timed words have \emph{strictly monotonic} timestamps,
which implies that some time (however small) must elapse between two consecutive transitions; other semantics 
have relaxed this requirement by allowing weakly monotonic timestamps  \cite{BY04}, thus permitting sequences of zero-time transitions. 
\end{itemize}

\emph{Hybrid automata} \cite{ACHH93,NOSY93,Hen96} are a generalization 
of timed automata where the dense-valued variables --- called 
``clocks'' in timed automata --- are permitted to evolve through 
more complicated timed behaviors. Namely, in hybrid automata 
one associates to each \emph{discrete state} a set of possible \emph{activities}, 
which are smooth functions (i.e., functions that are continuous 
together with all of their derivatives) from time to the dense 
domain of the \emph{variables}, and a set of \emph{invariants}, which 
are sets of allowed values for the variables. Activities specify 
possible variables' behaviors, thus generalizing the simple dynamics 
of clock variables in timed automata. More explicitly, whenever 
a hybrid automaton sits in some discrete location, its variables 
evolve over time according to one activity, \emph{nondeterministically} 
chosen among those associated with that state. However, the evolution 
can continue only as long as the variables keep their values 
within the invariant set of the state. Then, upon reading input 
symbols, the automaton instantaneously switches its discrete 
state, possibly resetting some variables according to the additional 
constraints attached to the taken transitions, similarly to timed 
automata.

\begin{figure}[htb!]
	 \centering
	 \includegraphics{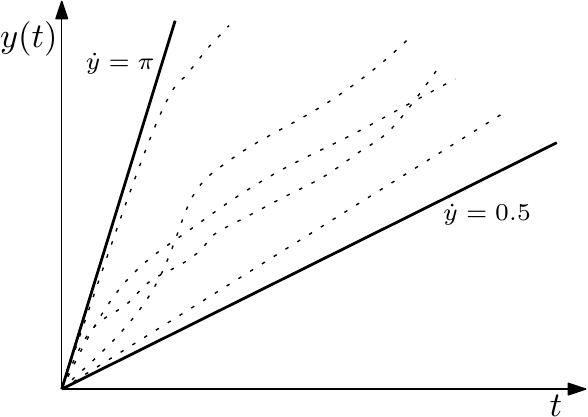}
	 \caption{Some behaviors compatible with the constraint $0.5 < \dot{y} < \pi$.}
	 \label{fig:hybrid_behavior}
\end{figure}

Although in this general definition the evolution of the dense-valued 
variables can be represented by any function such that all its derivatives are continuous, in practice 
more constrained (and simply definable) subsets are usually considered. 
A common choice is to define the activities by giving a set of 
bounds on the first-order derivative, with respect to time, of 
the variables. For a variable $y$, the constraint $0.5 < \dot{y} < \pi$
is an example of a class of such activities (see Figure \ref{fig:hybrid_behavior} for a visual
representation).

In both timed and hybrid automata, one typically defines a \kw{composition}
semantics where concurrent automata evolve in parallel, but synchronize 
on transitions in response to input symbols, similarly to traditional 
automata and Statecharts.

The development of timed and hybrid automata was also motivated 
by the desire to extend and generalize the powerful and
successful techniques of automatic \kw{verification} (and model checking 
in particular) based on the combination of infinite-word finite-state 
automata and temporal logic (see Section \ref{sec:duallanguage}), to the metric 
treatment of time. However, the presence of real-valued variables 
renders the verification problem much more difficult and, often, 
undecidable. Thus, with respect to the general model, restrictions 
are introduced that make the models more tractable and amenable 
to verification --- usually at the price of sacrificing some expressiveness.

In a nutshell, the verification problem is generally tackled 
by producing a \emph{finite abstraction} of a timed/hybrid automaton, 
where all the relevant behaviors of the modeled system are captured 
by an equivalent, but finite, model, which is therefore exhaustively 
analyzable by model checking techniques. Such procedures usually 
assume that all the numeric constraints on clocks and variables 
are expressed by \emph{rational} numbers; this permits the partitioning
of the space of all possible behaviors of the variables into a finite 
set of \emph{regions} that describe equivalent behaviors, preserving 
verification properties such as reachability and emptiness. 
For a precise description of these techniques see e.g., \cite{AM04,ACHHHNOSY95,HNSY94,HKPV98}.

These analysis techniques have been implemented in some interesting 
tools, such as UPPAAL \cite{LPY97}, Kronos \cite{Yov97}, Cospan \cite{AK95}, IF \cite{BGOOS04}, and HyTech \cite{HHW97}.

\paragraph{Timed Transition Models.}
Ostroff's \emph{Timed Transition Models} (TTM) \linebreak \cite{Ost90} are another 
formalism that is based on enriching automata with time variables; 
they are a real-time \kw{metric} extension of Manna and Pnueli's fair transition 
systems \cite{MP92}.

In TTMs, time is modeled \kw{explicitly} by means of a clock variable $t$.
$t$ takes values in a \kw{discrete} time domain, and is updated explicitly 
and \kw{synchronously} by the occurrence of a special \emph{tick} transition. The clock 
variable, as any variable in TTMs, is global and thus shared 
by all transitions. All transitions other than \emph{tick} do not 
change time but only update the other components of the state; 
therefore it is possible to have several different states associated 
with the same time instant. Transitions are usually annotated 
with lower and upper bounds $l, u$; this prescribes that 
the transition is taken at least $l$, and no more than $u$
clock ticks (i.e., time units), after the transition has become 
enabled.

\begin{figure}[htb!]
	 \centering
	 \includegraphics{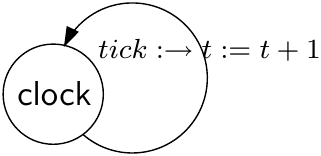}
	 \caption{A Timed Transition Model for the clock.}
	 \label{fig:tick}
\end{figure}

In practice, it is assumed that every TTM system includes a \emph{global clock}
subsystem, such as that pictured in Figure \ref{fig:tick}. Notice that 
this subsystem allows the special \emph{tick} transition to occur 
at any time, making time advance one step. The \emph{tick} transition 
is \emph{a priori} assumed to be fairly scheduled, that is it must 
occur infinitely often to prevent \kw{Zeno} behaviors where time stops. 

We give a few more details of TTMs in Section \ref{sec:duallanguage} (where a TTM 
resource manager specification is also given) when discussing 
dual language approaches.

\subsubsection{Asynchronous Abstract Machines: Petri nets} \label{sec:petrinets}
This section introduces Petri nets as one of the most popular 
examples of asynchronous abstract machines.

Petri nets owe their name to their inventor, Carl Adam Petri \cite{Pet63}.
Since their introduction they became rather popular 
both in the academic and, to some extent, in the industrial world, 
as a fairly intuitive graphical tool to model concurrent systems. 
For instance, they inspired transition diagrams adopted in the 
UML standard \cite{UML05,UML04,EPLF03}. There are a few slightly 
different definitions of such nets and of their semantics. Among 
them one of the most widely adopted is the following, which we 
present informally; the reader is referred to the literature \cite{Pet81,Rei85}
for a comprehensive treatment.

A \emph{Petri net} consists of a set of places, and a set of transitions. 
Places store tokens and pass them to transitions. A transition 
is \emph{enabled} whenever all of the incoming places hold at least 
one token. Whenever a transition is enabled a \emph{firing} can 
occur; this happens nondeterministically. As a consequence of 
a firing, the enabling tokens are removed from the incoming places 
and moved to the outgoing places the transition is connected 
to. Thus, for any possible combination of nondeterministic choices, 
we have a \emph{firing sequence}.

\begin{figure}[htb!]
	 \centering
	 \includegraphics{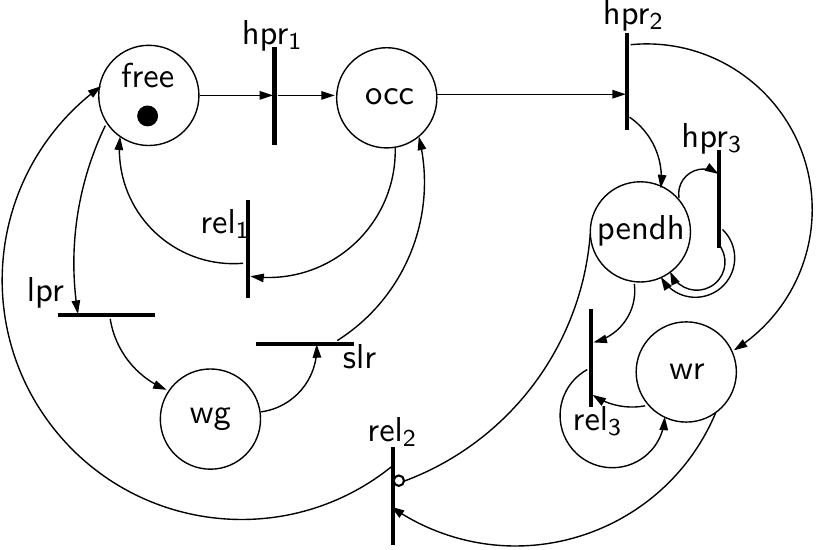}
	 \caption{A resource manager modeled through a Petri net.}
	 \label{fig:petri_net_untimed}
\end{figure}

Let us consider again the example of the resource manager, using 
a Petri net model. We introduce the following modifications with 
respect to the previous examples. First, since we are now considering 
untimed Petri nets, we do not introduce any metric time constraint. 
Second, we disallow low-priority requests while the resource 
is occupied, or high-priority requests while there is a pending 
low-priority request. Conversely, we introduce a mechanism to 
``count'' the number of consecutive high-priority requests that 
occur while the resource is occupied. Then, we make sure that 
all of them are served (consecutively) before the resource becomes 
free again. This behavior is modeled by the Petri net in Figure \ref{fig:petri_net_untimed}, 
where the places are denoted by the circles $\free$, $\occ$, $\pendh$, $\wrr$, 
and $\wg$, and the thick lines denote transitions. Notice that 
we allow an unbounded number of tokens in each place (actually, 
the only place where the tokens can accumulate is $\pendh$, 
where each token represents a pending high-priority request). 
Finally, we have also chosen to introduce an \emph{inhibiting arc}, 
from place $\pendh$ to transition $\rel_{2}$, denoted by a small circle 
in place of an arrowhead: this means that the corresponding transition 
is enabled if and only if place $\pendh$ stores no tokens. This 
is a non-standard feature of Petri nets which is often added 
in the literature to increase the model's expressive power.

According to our taxonomy, Petri nets, as defined above, can 
be classified as follows: 

\begin{itemize}
\item There is no explicit notion of time. However a time model can 
be \kw{implicitly} associated with the semantics of the net.

\item There are at least two major approaches to formalizing 
the semantics of Petri nets. 
  \begin{itemize}
  \item The simpler one is based on \emph{interleaving semantics}. According 
	 to this semantics the behaviors of a net are just its firing 
	 sequences. Interleaving semantics, however, introduces a total 
	 ordering in the events modeled by the firing of net transitions 
	 which fails to capture the asynchronous nature of the model. 
	 For instance, in the net of Figure \ref{fig:petri_net_untimed} the two sequences
	 $\langle \hpr_1, \hpr_2, \hpr_3, \rel_3, \hpr_3, \rel_3, \rel_3, \rel_2 \rangle$ and
	 $\langle \hpr_1, \hpr_2, \hpr_3, \hpr_3, \rel_3, \rel_3, \rel_3, \rel_2 \rangle$
	 both belong to the set of possible net's behaviors; however, 
	 they both imply an order between the firing of transitions $\hpr_3$ and $\rel_3$, 
	 whereas the graphical structure of the net emphasizes that the 
	 two events can occur asynchronously (or simultaneously).

  \item For this reason, a \emph{true concurrency} (i.e., fully \kw{asynchronous}) 
	 approach is often preferred to describe the semantics of Petri 
	 nets. In a true concurrency approach it is natural to see the 
	 time model as a \emph{partial order}, instead of a total order of 
	 the events modeled by transition firings. Intuitively, in a true 
	 concurrency modeling the two sequences above can be ``collapsed'' into
	 $\langle \hpr_1, \hpr_2, \hpr_3, \{ \hpr_3, \rel_3 \}, \rel_3, \rel_3, \linebreak \rel_2 \rangle$,
	 where the pair $\{\ \}$ denotes the fact that the included items can 
	 be ``shuffled'' in any order.
  \end{itemize}

\item Petri nets are a \kw{nondeterministic} operational model. For instance, 
still in the net of Figure \ref{fig:petri_net_untimed}, whenever place $\occ$ holds 
some tokens, both transitions $\hpr_2$ and $\rel_1$ are enabled, but they are 
in \emph{conflict}, so that only one of them can actually fire. 
Such a nondeterminism could be formalized by exploiting a \kw{branching}\emph{-time} model.

\item In traditional Petri nets the time model has no \kw{metrics}, so that 
it should be seen only as a (possibly partial) order.\footnote{Unless one adopts the convention of associating one time unit to the firing of a single transition, as it is often assumed in other --- synchronous --- operational models such as finite state automata. Such an assumption, however, would contrast sharply with the asynchronous original nature of the model.}

\item We also remark that Petri nets are usually ``less compositional'' 
than other operational formalisms, and synchronous automata in 
particular. While notions of \kw{composition} of Petri nets have been 
introduced in the literature, they are often less natural and 
more complicated than, for instance, Statecharts' synchronous 
composition; this is partly due to the asynchronous ``nature'' 
of the nets. 
\end{itemize}

To model hard real-time systems a metric time model is necessary in
most cases. To overcome this difficulty, many extensions have been
proposed in the literature to introduce a metric time model. Here we
report on Merlin and Farber's approach \cite{MF76}, which has been
probably the first one of such extensions and is one of the most
intuitive and popular ones.  For a thorough and comprehensive survey
of the many time extensions to Petri nets we refer to
\cite{CM99,Cer93}.

\begin{figure}[htb!]
	 \centering
	 \includegraphics{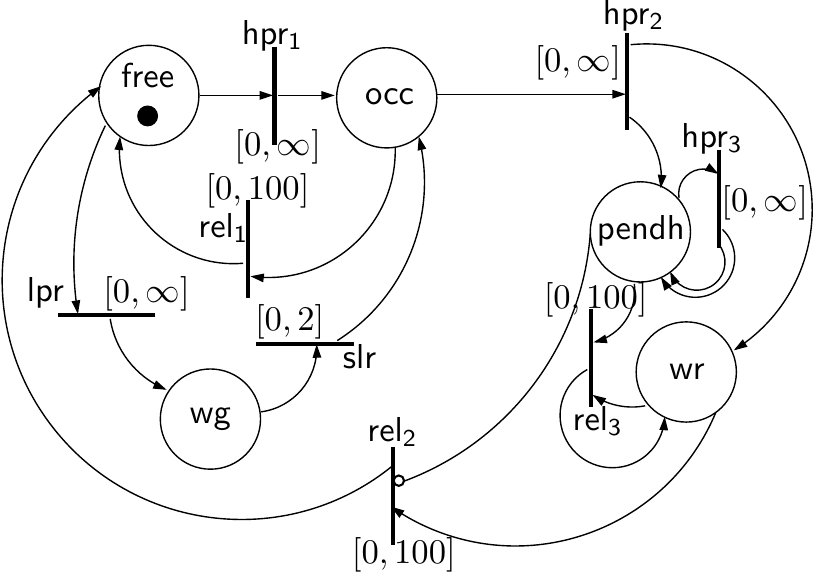}
	 \caption{A resource manager modeled through a timed Petri net.}
	 \label{fig:petri_net}
\end{figure}

A Timed Petri net according to the Merlin and Farber's approach is
simply a net where a minimum and a maximum firing time are attached to
each transition (both firing times can be 0, and the maximum time can
be $\infty$). Figure \ref{fig:petri_net} shows how the net of Figure
\ref{fig:petri_net_untimed} can be augmented in such a way. The time
bounds that have been introduced refine the specification of the
resource manager by prescribing that each use of the resource must
take no longer than 100 contiguous (i.e., since the last request
occurred) time units, and that a low priority request is served within
2 time units.

The fairly natural intuition behind this notation is that, since the
time when a transition is enabled (i.e., all its input places have
been filled with at least one token), the transition can fire ---
nondeterministically --- at any time that is included in the specified
interval, unless it is disabled by the firing of a conflicting
transition. For instance, place $\wg$ becomes occupied after a low
priority request is issued, thus enabling transition $\slr$. The
latter can fire at any time between 0 and 2 time instants after it has
become enabled, thus expressing the fact that the request is served
\emph{within} 2 time units.

Despite its intuitive attractiveness, several intricacies are 
hidden in the previous informal definition, as has been pointed 
out in the literature when attempting to formalize their semantics \cite{FMM94,GMMP91}. 
Here we focus only on the main ones.

\begin{itemize}
\item Suppose that the whole time interval elapsed since the time when 
a transition became enabled: is at this point the transition \emph{forced} 
to fire or not? In the negative case it will never fire in the 
future and the tokens in their input places will be wasted (at 
least for \emph{that} firing). There are arguments in favor of both 
choices; normally --- including the example of Figure \ref{fig:petri_net} --- the 
former one is assumed (it is often called \emph{strong time semantics} 
(STS)) but there are also cases where the latter one (which is 
called \emph{weak time semantics} (WTS) and is considered as more 
consistent with traditional Petri nets semantics, where a transition 
is never forced to fire) is preferred.\footnote{In this regard, notice that the timed automata of Section \ref{sec:synchronous} could be considered to have a \emph{weak time semantics}. In fact, transitions in timed automata \emph{are not forced} to be taken when the upper limit of some constraint is met; rather, all that is prescribed by their semantics is that \emph{when} (if) a transition is taken by a timed automaton, its corresponding constraint (and those of the source and target locations) \emph{must} be met.}

\item If the minimum time associated with a transition is 0, then the
  transition can fire immediately once enabled and we have a case of
  zero-time transition (more precisely we call this circumstance
  \emph{zero-time firing} of the transition). As we pointed out in
  other cases, zero-time firing can be a useful abstraction whenever
  the duration of the event modeled by the firing of the transition
  can be neglected with respect to other activities of the whole
  process.\footnote{Normally the firing of a transition is considered
    as instantaneous. This assumption does not affect generality since
    an activity with a non-null duration can be easily modeled as a
    pair of transitions with a place between them: the first
    transition models the beginning of the activity and the second one
    models its end.}  On the other hand zero-time firing can produce
  some intricate situations since two subsequent transitions (e.g.,
  $\hpr_1$ and $\rel_1$ in Figure \ref{fig:petri_net}) could fire
  simultaneously. This can produce some \kw{Zeno} behaviors if the net
  contains loops of transitions with 0 minimum time. For this reason
  ``zero-time loops'' are often forbidden in the construction of timed
  Petri nets.
\end{itemize}

Once the above semantic ambiguities have been clarified, the behavior 
of timed Petri nets can be formalized through two main approaches. 

\begin{itemize}
\item A time stamp can be attached to each token when it is produced 
by the firing of some transition in an output place. For instance, 
with reference to Figure \ref{fig:petri_net}, we might have the sequence of
transitions $\langle \hpr_1(2), \hpr_2(3), \linebreak \hpr_3(4), \rel_3(5), \hpr_3(6), \cdots \rangle$
(that is, $\hpr_1$ fires at time 2 producing a token with 
time stamp 2 in $\occ$; this is consumed at time 3 by the firing 
of $\hpr_2$ which also produces one token in $\pendh$ and 
one in $\wrr$, both timestamped 3, etc.). In this way time is 
\kw{explicitly} modeled in a \kw{metric} way --- whether \kw{discrete} or \kw{continuous} 
--- as a further variable describing system's state and evolution 
(more precisely, as \emph{many} further variables, one for each 
produced token).

As remarked in Section \ref{sec:synchronous}, this approach actually introduces 
two different time models in the formalism: the time implicitly 
subsumed by the firing sequence and the time modeled by the time 
stamps attached to tokens. Of course some restrictions should 
be applied to guarantee consistency between the two orderings: 
for instance, the same succession of firings described above could induce the timed 
sequence $\langle \hpr_1(2), \hpr_2(3), \hpr_3(4), \linebreak \hpr_3(6), \rel_3(5), \cdots \rangle$,
that should however be excluded from the possible behaviors. 

\item The net could be described as a dynamical system as in the traditional 
approach described in Section \ref{sec:dynamicalsys}. The system's state would be 
the net marking whose evolution should be formalized as a function 
of time. To pursue this approach, however, a few technical difficulties 
must be overcome:
  \begin{itemize}
  \item First, tokens cannot be formalized as entities with no identity, 
	 as it happens with traditional untimed Petri nets. Here too, 
	 some kind of time stamp may be necessary. Consider, for instance, 
	 the net fragment of Figure \ref{fig:petri_state}, and suppose that one token is 
	 produced into place $\mathsf{P}$ at time 3 by transition $\mathsf{t^i_1}$ and another 
	 token is produced by $\mathsf{t^i_2}$ at time 4; then, according to the normal 
	 interpretation of such Petri nets (but different semantic formalizations 
	 could also be given, depending on the phenomenon that one wants 
	 to model) the output transition $\mathsf{t^o}$ should fire once at time 6=3+3 
	 and a second time at time 7=4+3. Thus, a state description that 
	 simply asserts that at time 4 there are two tokens in $\mathsf{P}$ would 
	 not be sufficient to fully describe the future evolution of the 
	 net.
\begin{figure}[htb!]
	 \centering
	 \includegraphics{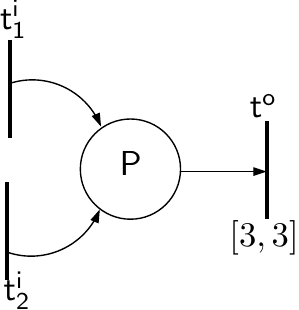}
	 \caption{An example Petri net fragment.}
	 \label{fig:petri_state}
\end{figure}

  \item If zero-time firings are admitted, strictly speaking, system's 
	 state cannot be formalized as a function of the independent variable 
	 ``time'': consider, for example, the case in which, in the net 
	 of Figure \ref{fig:petri_net}, at time $t$ both transitions $\lpr$ and $\slr$ fire
	 (which can happen, since $\slr$ admits zero-time firing); in this case, it 
	 would happen that at time $t$ both a state (marking) where 
	 place $\wg$ is marked and a state where place $\occ$ is marked 
	 --- and $\wg$ is not marked anymore --- would hold.

	 In \cite{FMM94} this problem has been solved by forbidding ``zero-time 
	 loops'' and by stating the convention that in case of a ``race'' 
	 of zero-time firings (which is always finite) only the places 
	 at the ``end of the race'' are considered as marked, whereas tokens 
	 flow instantaneously through other places without marking them. 

	 In \cite{GMM99} a more general approach is proposed, where zero-time 
	 firings are considered as an abstraction of a non-null but \emph{infinitesimal} 
	 firing time. By this way it has been shown that mathematical 
	 formalization and analysis of the net behavior become simpler 
	 and --- perhaps --- more elegant.
    \end{itemize}
\end{itemize}

Timed Petri nets have also been the object of a formalization through 
the dual language approach (see Section \ref{sec:duallanguage}).

As for other formalisms of comparable expressive power, Petri nets
suffer intrinsic limitations in the techniques for (semi-)automatic
\kw{analysis} and \kw{verification}. In fact, let us consider the reachability
problem, i.e., the problem of stating whether a given marking can be
reached by another given marking. This is the main analysis problem
for Petri nets since most other properties can be reduced to some
formulation of this basic problem \cite{Pet81}. For normal, untimed
Petri nets with no inhibitor arcs the reachability problem has been
shown intractable though decidable; if Petri nets are augmented with
some metric time model and/or inhibitor arcs, then they reach the
expressive power of Turing machines and all problems of practical
interest become undecidable.\footnote{Of course, interesting
  particular cases are always possible, e.g., the case of bounded
  nets, where the net is such that during its behavior the number of
  tokens in every place never exceeds a given bound.} Even building
interpreters for Petri nets to analyze their properties through
simulation faces problems of combinatorial explosion due to the
intrinsic nondeterminism of the model.

Nevertheless interesting tools for the analysis of both untimed and
timed Petri nets are available. Among them we mention \cite{BD91},
which provides an algorithm for the reachability problem of timed
Petri nets assuming the set of rational numbers as the time
domain. This work has pioneered further developments. For a
comprehensive survey of tools based on Petri nets see \cite{TGI}.

\paragraph{}
Before closing this section let us also mention the Abstract State
Machines (ASM) formalism \cite{BS03}, whose generality subsumes most
types of operational formalisms, whether synchronous or
asynchronous. However, ASM have not received, to the best of our
knowledge, much attention in the realm of real-time computing until
recently, when the Timed Abstract Machine notation \cite{OL07} and
its tools \cite{OL07b} have been developed.

\subsection{Descriptive Formalisms} \label{sec:descriptive} Let us now
consider \emph{descriptive} (or \emph{declarative}) formalisms.  In
descriptive formalisms a system is formalized by declaring the
fundamental properties of its behavior. Most often, this is done by
means of a language based on mathematical logic; more seldom algebraic
formalisms (e.g., process algebras) are exploited. As we saw in
Section \ref{sec:metamodel}, descriptive notations can be used alone
or in combination with operational ones, in a dual language
approach. In the former case, both the requirements and the system
specification are expressed within the same formalism; therefore
verification consists of proving that the axioms (often expressed in
some logic language) that constitute the system specification imply
the formulas that describe the requirements. In the latter case, the
verification is usually based on some \emph{ad hoc} techniques, whose
features may vary significantly depending on the adopted combination
of descriptive and operational notations. We treat dual-language
approaches in Section \ref{sec:duallanguage}.

When considering the description of the timed behavior of a system 
through a logic formalism, it is natural to refer to \emph{temporal 
logics}. A distinction should be made here. Strictly speaking, 
temporal logics are a particular family of modal logics \cite{Kri63,RU71} 
possessing specific operators --- called \emph{modalities} --- apt 
to express temporal relationships about time-dependent propositions. 
The modalities usually make the treating of time-related information 
quite intuitive as they avoid the \kw{explicit} reference to absolute 
time values and mirror the way the human mind intuitively reasons 
about time; indeed, temporal logics were initially introduced 
by philosophers \cite{Kam68}. It was Pnueli who first observed \cite{Pnu77} 
that they could be effectively used to reason about temporal 
properties of programs, as well. Some temporal logics are discussed 
in the following Section \ref{sec:temporallogics}.

In the computer science communities, however, the term ``temporal 
logic'' has been used in a broader sense, encompassing all logic-based 
formalisms that possess some mechanism to express temporal properties 
and to reason about time, even when they introduce some \kw{explicit} 
reference to a dedicated variable representing the current value 
of time or some sort of clock and hence adopt a style of description 
that is different in nature from the original temporal logic 
derived from modal logic. Many of these languages have been used 
quite successfully for modeling time-related features: some of 
them are described in Section \ref{sec:logicswtime} below.

We emphasize that there is a wide variety of different styles 
and flavors when it comes to temporal logics. As usual, we do 
not aim to be exhaustive in the presentation of temporal logics 
(we refer the reader to other papers specifically on temporal 
logics, e.g., \cite{Eme90,AH93,AH92,Ost92,Hen98,BMN00,FPR}), but 
to highlight some significant approaches to the problem of modeling 
time in logic.

Finally, a different approach to descriptive modeling of systems, 
based on the calculational aspects of specifications, is the 
algebraic one. We discuss algebraic formalisms in Section \ref{sec:algebraic}.

\subsubsection{Temporal Logics} \label{sec:temporallogics} In this
section we deal with temporal logics with essentially \kw{implicit} time,
and we focus our discussion on a few key issues, namely the
distinction between linear-time and branching-time logics, the
adoption of a discrete or non-discrete time model, the use of a metric
on time to provide means to express temporal properties in a
quantitatively precise way, the choice of using solely temporal
operators that refer to the future versus introducing also past-tense
operators, and the assumption of time points or time intervals as the
fundamental time entities. In our discussion we will go from simple to
richer notations and occasionally combine the treatment of some of the
above mentioned issues.  Finally, some \kw{verification} issues about
temporal logics will be discussed while presenting dual language
approaches in Section \ref{sec:duallanguage}.

\paragraph{Linear-Time Temporal Logic.}
As a first, simplest example of temporal logic, let us consider 
propositional Linear-Time Temporal Logic (LTL) with discrete 
time. In LTL, formulas are composed from the atomic propositions 
with the usual Boolean connectives and the temporal connectives $\XLTL$ 
(\emph{next}, also denoted with the symbol $\bigcirc$), $\FLTL$ (\emph{eventually 
in the future}, also $\Diamond$), $\GLTL$ (\emph{globally} --- i.e., \emph{always} 
--- \emph{in the future}, also $\Box$), and $\ULTL$ (\emph{until}). These 
have a rather natural and intuitive interpretation, as the formulas 
of LTL are interpreted over \kw{linear} sequences of states: the formula $\XLTL \pp$ 
means that proposition $\pp$ holds at the state that immediately 
follows the one where the formula is interpreted, $\FLTL \pp$ 
means that $\pp$ will hold at some state following the current 
one, $\GLTL \pp$ that $\pp$ will hold at all future states, $\pp \ULTL \qq$ 
means that there is some successive state such that proposition $\qq$ 
will hold then, and that $\pp$ holds in all the states between 
the current and that one.

Notice that the presence of the ``next'' operator $\XLTL$ implies 
that the logic refers to a \kw{discrete} temporal domain: by definition, 
there would be no ``next state'' if the interpretation structure 
domain were not discrete. On the other hand, depriving LTL of 
the next operator would ``weaken'' the logic to a pure ordering 
without any metrics (see below).

To illustrate LTL's main features, let us consider again the 
resource manager introduced in the previous sections: the following 
formula specifies that, if a low priority request is issued at 
a time when the resource is free, then it will be granted at 
the second successive state in the sequence. 
\begin{equation*}
  \GLTL ( \mathsf{free} \wedge \mathsf{lpr}  \Rightarrow \XLTL \XLTL \mathsf{occ} )
\end{equation*}

LTL is well-suited to specify qualitative time relations, for 
instance ordering among events: the following formula describes 
a possible assumption about incoming resource requests, i.e., 
that no two consecutive high priority requests may occur without 
a release of the resource between them (literally, the formula 
reads as: if a high priority request is issued then the resource must be eventually released and no other 
similar request can take place until the release occurs). 
\begin{equation*}
  \GLTL ( \mathsf{hpr} \Rightarrow \XLTL ( \neg \mathsf{hpr} \, \ULTL \,\mathsf{rel} ))
\end{equation*}

Though LTL is not expressly equipped with a \kw{metric} on time, one might
use the next operator $\XLTL$ for this purpose: for instance, $\XLTL^3
\pp$ (i.e., $\XLTL \XLTL \XLTL \pp$) would mean that proposition $\pp$
holds 3 time units in the future. The use of $\XLTL^k$ to denote the
time instant at $k$ time units in the future is only possible,
however, under the condition that there is a one-to-one correspondence
between the states of the sequence over which the formulas are
interpreted and the time points of the temporal domain. Designers of
time-critical systems should be aware that this is not necessarily the
case: there are linear discrete-time temporal logics where two
consecutive states may well refer to the same time instant whereas the
first following state associated with the successive time instant is
far away in the state sequence \cite{Lam94,MP92,Ost89}. We already
encountered this critical issue in the context of finite state
automata and the \kw{fairness} problem (see Section \ref{sec:synchronous})
and timed Petri nets when zero-time transitions are allowed (see
Section \ref{sec:petrinets}) and will encounter it again in the dual
language approach (Section \ref{sec:duallanguage}).

\paragraph{Metric Temporal Logics.}
Several variations or extensions of \kw{linear} time temporal logic have
been defined to endow it with a metric on time, and hence make it
suitable to describe strict real-time systems. Among them, we mention
Metric Temporal Logic (MTL) \cite{Koy90} and TRIO \cite{GMM90,MMG92}.
They are commonly interpreted both over \kw{discrete} and over \kw{dense} (and
\kw{continuous}) time domains.

MTL extends LTL by adding to its operators a \kw{quantitative} time
parameter, possibly qualified with a relational symbol to imply an
upper bound for a value that typically represents a distance between
time instants or the length of some time interval.  For instance the
following simple MTL formula specifies bounded response time: there is
a time distance $d$ such that an event $\pp$ is always followed by an
event $\qq$ with a delay of at most $d$ time units (notice that MTL is
a first-order logic).
\begin{equation} \label{eq:MTL}
  \exists d: \GLTL ( \pp \Rightarrow \FLTL_{< d} \,\qq)
\end{equation}

The following formula asserts that $\pp$ eventually takes place, and then periodically occurs with period $d$.
\begin{equation*}
  \FLTL ( \pp \wedge \GLTL (\neg \pp \, \ULTL_d \, \pp) )
\end{equation*}

TRIO introduces a quantitative notion of time by adopting a single 
basic modal operator, called \emph{Dist}. The simplest formula $\Dist{\pp, d}$
means that proposition $\pp$ holds at a time instant exactly $d$
time units from the current one; notice that this formula may 
refer to the future, if $d > 0$, or to the past, if $d < 0$, 
or even to the present time if $d = 0$. All the operators of 
LTL, their quantitative-time counterparts and also other operators 
not found in traditional temporal logic are defined in TRIO by 
means of first-order quantification over the time parameter of 
the basic operator \emph{Dist}. We include in Table \ref{tab:trio-operators} a list of some 
of the most significant ones (and especially those used in the following).

\begin{table}[tbh]
\begin{center}
  \begin{scriptsize}
  \begin{tabular}{|c|c|p{4.8cm}|}
    \hline
    \textsc{Operator}        &  \textsc{Definition}  &   \textsc{Description}\\
    \hline
    $\Futr{F,t}$             &  $t \geq 0 \wedge \Dist{F,t}$   &   $F$ holds $t$ time units in the future \\
    $\Past{F,t}$             &  $t \geq 0 \wedge \Dist{F,-t}$  &   $F$ held $t$ time units in the past \\

    $\Alw{F}$                 &  $\forall d: \Dist{F,d}$       &   $F$ holds always \\

    $\Lasts{}{F,t}$            &  $\forall d \in (0,t): \Futr{F,d}$ &  $F$ holds for $t$ time units in the future \\
    $\Lasted{}{F,t}$           &  $\forall d \in (0,t): \Past{F,d}$ &  $F$ held for $t$ time units in the past \\

    $\WithinF{}{F,t}$          &  $\exists d \in (0,t): \Futr{F,d}$ &  $F$ holds within $t$ time units in the future \\

   $\Until{F,G}$            &  $\exists d > 0: \Lasts{}{F,d} \wedge \Futr{G,d}$  &  $F$ holds until $G$ holds \\ 

    $\NowOn{F}$             &  $\exists d > 0: \Lasts{}{F,d}$ &   $F$ holds for some non-empty interval in the future \\

   $\UpToNow{F}$             &  $\exists d > 0: \Lasted{}{F,d}$ &   $F$ held for some non-empty interval in the past \\
    \hline
  \end{tabular}
  \end{scriptsize}
\end{center}
\caption{TRIO derived temporal operators.}
\label{tab:trio-operators}
\end{table}

Referring again to the example of the resource manager, the following 
TRIO formula asserts that any low priority resource request is 
satisfied within 100 time units 
\begin{equation*}
  \Alw{\mathsf{lpr} \Rightarrow \WithinF{}{\mathsf{occ}, 100}}
\end{equation*}
while the next one states that any two high priority requests 
must be at least 50 time units apart. 
\begin{equation*}
  \Alw{\mathsf{hpr} \Rightarrow \Lasts{}{\neg \mathsf{hpr}, 50}}
\end{equation*}

We note incidentally that both in MTL and in TRIO the interpretation 
structure associates one single state with every time instant 
and no \kw{explicit} state component needs to be devoted to the representation 
of the current value of ``time'': quantitative timing properties 
can be specified using the modal operators embedded in the language. 
Other approaches to the quantitative specification of timing 
properties in real-time systems are based on the use of the operators 
of (plain) LTL in combination with assertions that refer to the 
value of some \emph{ad hoc} introduced clock predicates or explicit 
time variable \cite{Ost89}. For instance the following formula of 
Real Time Temporal Logic (RTTL, a logic that will be discussed 
in Section \ref{sec:duallanguage}) states the same property expressed by 
MTL Formula (\ref{eq:MTL}) above specifying bounded response time (in 
the formula the variable $t$ represents the current value of 
the time state component). 
\begin{equation*}
  \forall \mathrm{T} \left( ( \pp \wedge t = \mathrm{T}) \Rightarrow \FLTL ( \qq \wedge t \leq \mathrm{T} + d ) \right)
\end{equation*}

\begin{figure}[htb!]
	 \centering
	 \includegraphics{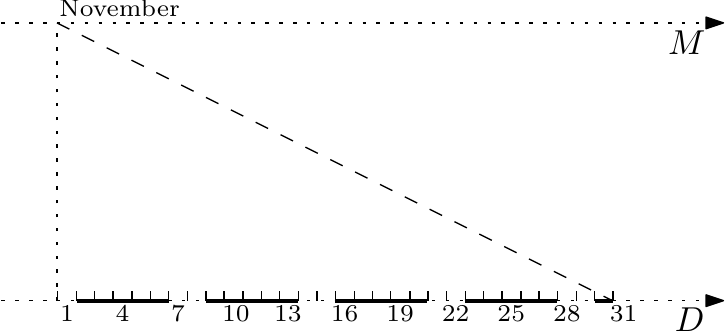}
	 \caption{Interpretation of an upper-level predicate in the lower-level domain. Solid lines denote the intervals in the lower domain where the predicate holds.}
	 \label{fig:granularity}
\end{figure}

\paragraph{Dealing with different time granularities}
Once suitable constructs are available to denote in a quantitatively
precise way the time distance among events and the length of time
intervals, then the problem may arise of describing systems that
include several components that evolve, possibly in a partially
independent fashion, on different time scales. This is dealt in the
temporal logic TRIO described above by adopting syntactic and semantic
mechanisms that enable dealing with different levels of \kw{time granularity} \cite{CCM+91}.
 Syntactically, temporal expressions can be
labeled in such a way that they may be interpreted in different time
domains: for instance, $30_D$ denotes 30 days whereas $3_H$ denotes 3
hours. They key issue is the possibility given to the user to specify
a semantic mapping between time domains of different granularity;
hence, the truth of a predicate at a given time value at higher
(coarser) level of granularity is defined in terms of the
interpretation in an interval at the lower (finer) level associated
with the value at the higher level. For instance, Figure
\ref{fig:granularity} specifies that, say, working during the month of
November means working from the 2$^\text{nd}$ through the 6$^\text{th}$, from the 9$^\text{th}$
through the 13$^\text{th}$, etc.

As with derived TRIO temporal operators, suitable predefined mappings
help the user specify a few standard situations. For instance given
two temporal domains $\timedomain_1$ and $\timedomain_2$, such that
$\timedomain_1$ is coarser than $\timedomain_2$, $\pp \eventin
\timedomain_1 \rightarrow \timedomain_2$ means that predicate $\pp$ is
true in any $t \in \timedomain_1$ if and only if it is true in just
one instant of the interval of $\timedomain_2$ corresponding to
$t$. Similarly, $\pp \completein \timedomain_1 \rightarrow
\timedomain_2$ means that $\pp$ is true in any $t \in \timedomain_1$ if and only if
it is true in the whole corresponding interval of $\timedomain_2$.

By this way the following TRIO formula
\begin{equation*}
\Alwgran{M}{\forall emp (\mathsf{work}(emp) \Rightarrow \mathsf{get\_salary}(emp))}
\end{equation*}
which formalizes the sentence ``every month, if an employee works,
then she gets her salary'' introduced in Section \ref{sec:discrete} is
given a precise semantics by introducing the mapping of Figure
\ref{fig:granularity} for predicate $\mathsf{work}$, and by stating
that $\mathsf{get\_salary} \eventin M \rightarrow D$.

In some applicative domains having administrative, business, or
financial implications, the change of time granularity is often paired
with a reference to a global time calendar that evolves in a
\emph{synchronous way}. For instance, time units such as days, weeks,
months and years change in a synchronized way at certain predefined
time instants (e.g., midnight or new year) that are conventionally
established in a global fashion.

On the contrary, when a process evolves in a way such that its
composing events are related directly with one another but are
unrelated with any global time scale, time distances can be expressed
in a time scale with no intended reference to a global time scale: in
such cases we say that time granularity is managed in an
\emph{asynchronous way}. Quite often the distinction of the two
intended meanings is implicit in natural language sentences and
depends on some conventional knowledge that is shared among the
parties involved in the described process; thus, in the formalization
stage, it needs to be made explicit.

Consider for instance the following description of a procedure for
carrying out written exams: ``Once the teacher has completed the
explanation of the exercise, the students must solve it within exactly
three hours. Then, the teacher will collect their solutions and will
publish and register the grades after three days''. Clearly, the
former part of the sentence must be interpreted in the asynchronous
way (students have to complete their job within 180 minutes starting
from the minute when the explanation ended). The latter part, however,
is normally intended according to the synchronous interpretation:
results will be published before midnight of the third ``calendar
day'' following the one when the exam was held.

This notion of synchronous vs. asynchronous refinement of predicates
can be made explicit by adding an indication ($S$ for
synchronous, $A$ for asynchronous) denoting the intended mode
of granularity refinement for the predicates included in the
subformula. Hence the above description of the written examination
procedure could be formalized by the following formula, where
$H$ stands for ``hours'', and $D$ for ``days'':

\begin{gather*}
\Alwgran{H,A}{\mathsf{exerciseDelivery} \Rightarrow \Futr{\mathsf{solutionCollect, 3}}} \\
\wedge \\
\Alwgran{D,S}{\mathsf{exerciseDelivery} \Rightarrow \Futr{\mathsf{gradesPublication, 3}}}
\end{gather*}

To the best of our knowledge only few other languages in the
literature approach the granularity problem in a formal way
\cite{BB06,Rom90}. Among these \cite{Rom90} addresses the problem both
for space and time in formal models of geographic data processing
requirements.

\paragraph{Dense Time Domains and the Non-Zenoness Property.}
The adoption of a dense, possibly continuous time domain allows 
one to model asynchronous systems where the occurrence of distinct, 
independent events may be at time instants that are arbitrarily 
close. As a consequence, \kw{Zeno} behaviors, where for instance an 
unbounded number of events takes place in a bounded time interval, 
become possible and must be ruled out by means of suitable axioms 
or through the adoption of \emph{ad hoc} underlying semantic assumptions. 
The axiomatic description of non-Zenoness is immediate for a 
first order, metric temporal logic like MTL or TRIO, when it is applied 
to simple entities like predicates or variables ranging over 
finite domains. It can be more complicated when non-Zenoness 
must be specified in the most general case of variables that 
are real-valued functions of time \cite{GM01}.

Informally, a predicate is non-Zeno if it has finite variability,
i.e., its truth value changes a finite number of times over any finite
interval. Correspondingly, a general predicate $\PP$ can be
constrained to be non-Zeno by requiring that there always exists a
time interval before or after every time instant, where $\PP$ is
constantly true or it is constantly false. This constraint can be
expressed by the following TRIO formula (see \cite{HR04,LWW07} for
formulations in other similar logics):
\begin{equation} \label{eq:nonZeno-pred}
  \Alw{ \left( \UpToNow{\PP} \vee \UpToNow{\neg \PP}\right) \wedge \left( \NowOn{\PP} \vee \NowOn{\neg \PP} \right)}
\end{equation}

The additional notion of non-Zeno \emph{interval-based} predicate 
is introduced to model a property or state that holds continuously 
over time intervals of length strictly greater than zero. Suppose, 
for instance, that the ``occupied state'' for the resource in 
the resource manager example is modeled in the specification 
through a predicate $\occ$; to impose that $\occ$ be an interval-based 
(non-Zeno) predicate, one can introduce, in addition to Formula (\ref{eq:nonZeno-pred}),
the following TRIO axiom (which eliminates the possibility 
of $\occ$ being true in isolated time instants). 
\begin{multline*}
  \Alw{} ( \left( \mathsf{occ} \Rightarrow \UpToNow{\mathsf{occ}} \vee \NowOn{\mathsf{occ}} \right) \wedge \\ %
	 \left( \neg \mathsf{occ} \Rightarrow \UpToNow{\neg \mathsf{occ}} \vee \NowOn{\neg \mathsf{occ}} \right) )
\end{multline*}

A complementary category of non-Zeno predicates corresponds to 
properties that hold at \emph{isolated time points}, and therefore 
can naturally model instantaneous events. If, in the resource 
manager specification, predicate $\hpr$ represents the issue 
of a high priority request, it can be constrained to be a point-based 
predicate by introducing the following formula in addition to 
Axiom (\ref{eq:nonZeno-pred}). 
\begin{equation*}
  \Alw{\UpToNow{\neg \mathsf{hpr}} \wedge \NowOn{\neg \mathsf{hpr}}}
\end{equation*}

Finally, non-Zenoness for a time dependent variable $T$ (representing
for instance the current temperature in a thermostat application)
ranging over an uncountable domain $D$ essentially coincides with $T$
being piecewise analytic,\footnote{A function is analytic at a given
  point if it possesses derivatives of all orders and agrees with its
  Taylor series about that point \cite{WeiA,Kno96}. It is piecewise
  analytic if it is analytic over finitely many contiguous (open)
  intervals.} as a function of time. Analyticity is a quite strong
``smoothness'' requirement on functions which guarantees that the
function intersects any constant line only finitely many times over
any finite interval. Hence, any formula of the kind $T = \mathrm{v}$,
where $\mathrm{v}$ is a constant value in $D$, is guaranteed to be
non-Zeno according to the previous definitions for
predicates. Formally, non-Zenoness for $T$ can be constrained by the
following TRIO formula (where $\mathsf{r}, \mathsf{l}: \reals \rightarrow D$ are functions
that are analytic at 0).
\begin{equation*}
  \Alw{\exists d > 0: \forall t: 0 < t < d \Rightarrow \left( \Dist{T = \mathsf{r}(t), t} \wedge \Dist{T = \mathsf{l}(t), -t} \right)}
\end{equation*} 

In \cite{GM06} it is shown that the adoption of a small set of predefined 
categories of specification items like the point- and interval-based 
predicates outlined above can make the modeling of real-time 
hybrid systems quite systematic and amenable to automated verification.

\paragraph{Future and Past Operators.}
While the Linear Temporal Logic LTL, as originally proposed by 
Pnueli \cite{Pnu77} to study the correctness of programs, has only 
future operators, one may consider additional modalities for the 
past tense, e.g., $\PLTL$ (for \emph{previous}) as the operator corresponding 
in the past to the next operator $\XLTL$, or $\OLTL$ (for \emph{once}) 
as opposed to $\FLTL$, $\SLTL$ (for \emph{since}) as the past version 
of the \emph{until} operator $\ULTL$, etc. The question then arises, 
whether the past operators are at all necessary (i.e., if they 
actually increase the expressiveness of the logic) or useful 
in practice (i.e., if there are significant classes of properties 
that can be described in a more concise and transparent way by 
using also past operators than by using future operators only).

Concerning the question of expressiveness, it is well known from 
\cite{GPSS80} that LTL with past operators does not add expressive 
power to future-only LTL. Moreover, the separation theorem by 
Gabbay \cite{Gab87} allows for the elimination of past operators, 
producing an LTL formula to be evaluated in the initial instant 
only: therefore, LTL with past operators is said to be \emph{initially equivalent}
to future-only LTL \cite{Eme90}.\footnote{As it is customary in the literature, we consider one-sided infinite time discrete domains (i.e., $\naturals$). The bi-infinite case (i.e., $\integers$) is much less studied \cite{PP04}.}

On the other hand, it is widely recognized that the extension 
of LTL with past operators \cite{Kam68} allows one to write specifications 
that are easier, shorter, and more intuitive \cite{LPZ85}. A customary 
example, taken from \cite{Sch02}, is the specification: \emph{Every alarm is due to a fault},
which, using the \emph{globally} operator $\GLTL$ and the \emph{previously} operator \emph{O}
(\emph{once}), may be very simply written as: 
\begin{equation*}
  \GLTL ( \mathsf{alarm} \Rightarrow \OLTL\, \mathsf{fault} )
\end{equation*}
whereas the following is one of the simplest LTL versions of 
the same specification, using the \emph{until} operator. 
\begin{equation*}
  \neg ( \neg \mathsf{fault} \,\ULTL\, (\mathsf{alarm} \wedge \neg \mathsf{fault} ))
\end{equation*}

In \cite{LMS02}, it has been shown that the elimination of past operators 
may yield an exponential growth of the length of the derived 
formula.

These expressiveness results change significantly when we consider 
logics interpreted over dense time domains. In general, past 
operators add expressive power when the time domain is dense, 
even if we consider mono-infinite time lines such as $\reals_{\geq 0}$. For instance, 
\cite{BCM05} shows that, over the reals, propositional MTL with past 
operators is strictly more expressive than its future-only version. 
The question of the expressiveness of past operators over dense 
time domains was first addressed, and shown to differ from the 
discrete case, in \cite{AH92b,AH93}.

\paragraph{Branching-Time Temporal Logic.}
As discussed in Section \ref{sec:linear}, in \kw{branching}\emph{-time temporal logic} 
every time instant may split into several future ones and therefore 
formulas are interpreted over \emph{trees} of states; such trees 
represent all possible computations of the modeled system. The 
branching in the interpretation structure naturally represents 
the \kw{nondeterministic} nature of the model, which may derive from 
some intrinsic feature of the device under construction or from 
some feature of the stimuli coming from the environment with 
which the device interacts. When interpreting a branching temporal 
logic formula at some current time, the properties asserted for 
the future may be evaluated with reference to \emph{all} future 
computations (i.e., branches of the state tree) starting from 
the current time or only to \emph{some} of them. Therefore, branching 
time temporal logic possesses modal operators that allow one 
to quantify universally or existentially over computations starting 
from the current time.

The Computation Tree Logic (CTL) \cite{EH86} has operators that are 
similar to LTL, except that every temporal connective must be 
preceded by a \emph{path quantifier}: either $\EXCTL$ (which stands 
for \emph{there exists a computation}, sometimes also denoted with 
the quantification symbol $\exists$) or $\ACTL$ (\emph{for all computations}, 
also $\forall$). With reference to the usual resource manager 
example, the formula below asserts that in every execution 
a low priority request (predicate $\lpr$) will be eventually followed 
by the resource being occupied (predicate $\occ$) in some of 
the evolutions following the request:
\begin{equation*}
  \ACTL \GLTL \left( \mathsf{lpr} \Rightarrow \EXCTL \FLTL \, \mathsf{occ} \right)
\end{equation*}
while the following formula asserts that there exists a computation 
of the resource manager where all low priority requests are certainly 
(i.e., in every possible successive evolution) eventually followed by the 
resource being occupied: 
\begin{equation*}
  \EXCTL \GLTL  \left( \mathsf{lpr} \Rightarrow \ACTL \FLTL \, \mathsf{occ} \right)
\end{equation*}

These examples, though very simple, show that in branching time
temporal logics temporal and path quantifiers may interact in quite a
subtle way.

Not surprisingly, branching temporal logic has been extended 
in a \kw{metric} version (TCTL, timed CTL) by adding to its operators 
quantitative time parameters, much in the same way MTL extends 
Linear Temporal Logic \cite{ACD93,HNSY94}. 

We refer the reader to \cite{Var01} for a deep analysis of the mutual 
pros and cons of linear time versus branching time logics.

\paragraph{Interval-Based Temporal Logics.}
All temporal logics we have considered so far adopt time \emph{points} 
as the fundamental entities: every state is associated with a 
time instant and formulas are interpreted with reference to some 
time instant. By contrast, the so-called \emph{interval temporal 
logics} assume time \emph{intervals}, rather than time instants, 
as the original temporal entity, while time points, if not completely 
ignored, are considered as derived entities. 

In principle, from a purely conceptual viewpoint, choosing intervals 
rather than points as the elementary time notion may be considered 
as a matter of subjective preference, once it is acknowledged 
that an interval may be considered as a set of points, while, 
on the other hand, a point could be viewed as a special case 
of interval having null length \cite{Koy92}. In formal logic, however, 
apparently limited variations in the set of operators may make 
a surprisingly significant difference in terms of expressiveness 
and complexity or decidability of the problems related with analysis 
and verification. Over the years, interval temporal logics have 
been a quite rich research field, producing a mass of formal 
notations with related analysis and verification procedures and 
tools.

A few relevant ones are: the Interval-based Temporal Logic of 
Schwartz et al.~\cite{SMV83}, the Interval Temporal Logic of Moszkowski \cite{Mos83,Mos86},
the Duration Calculus of Chachoen et al.~\cite{CHR91}, the Metric Interval Temporal Logic (MITL)
of Alur et al.~\cite{AFH96}. 

Among them, Duration Calculus (DC) refers to a \kw{continuous linear} sequence of
time instants as the basic interpretation structure. The significant
portions of the system state are modeled by means of suitable
functions from time (i.e., from the nonnegative reals) to Boolean
values, and operators measuring accumulated durations of states are
used to provide a \kw{metric} over time. For instance, in our resource
manager example, the property that the resource is never occupied for
more than 100 time units without interruption (except possibly for
isolated instants) would be expressed with the DC formula:
\begin{equation*}
  \Box ( \lceil \occ \rceil \Rightarrow \DClen \leq 100 )
\end{equation*}
where $\lceil \occ \rceil$ is a shorthand for $\int \occ = \DClen \wedge \DClen > 0$, which formalizes the 
fact that the predicate $\occ$ stays true continually (except for isolated points) over an interval of length $\DClen$.

Another basic operator of Duration Calculus (and of several other 
interval logics as well) is the \emph{chop} operator ; (sometimes denoted as $^\cap$).
Its purpose it to join two formulas predicating 
about two different intervals into one predicating about two 
adjacent intervals. For example, if we wanted to formalize 
the property that any client occupies the resource for at least 
5 time units, we could use the chop operator as follows:
\begin{equation*}
  \Box ( \lceil \neg \occ \rceil ; \lceil \occ \rceil ; \lceil \neg \occ \rceil \Rightarrow \DClen > 5 )
\end{equation*}
where the symbol $\DClen$ in the right-hand side of the implication 
now refers to the length of the overall interval, obtained by 
composition through the \emph{chop} operator.

Duration Calculus also embeds an underlying semantic assumption 
of finite variability for state functions that essentially corresponds 
to the previously discussed non-\kw{Zeno} requirement: each (Boolean-valued) 
interpretation must have only finitely many discontinuity points 
in any finite interval.

\subsubsection{Explicit-Time Logics} \label{sec:logicswtime}
Another category of descriptive formalisms adopts a ``timestamp'' 
\kw{explicit} view of time. This is typically done by introducing an \emph{ad hoc}
feature (e.g., a variable that represents the current time, 
or a time-valued function providing a timestamp associated with 
every event occurrence). In this section we focus on the distinguishing 
features of Lamport's Temporal Logic of Actions (TLA) \cite{Lam94}, 
and Alur and Henzinger's Timed Propositional Temporal Logic (TPTL) \cite{AH94}.
Other relevant examples of explicit-time logics are the Real Time Logic (RTL) of Mok et al.~\cite{JM86}
and Ostroff's Real-Time Temporal Logic (ESM/RTTL) \cite{Ost89} (which will be presented in 
the context of the dual language approach in Section \ref{sec:duallanguage}).

\paragraph{Temporal Logic of Actions.}
TLA formulas are interpreted over \kw{linear}, \kw{discrete} state sequences, 
and include variables, first order quantification, predicates 
and the usual modal operators $\FLTL$ and $\GLTL$ to refer to some 
or all future states. While basic TLA does not have a \kw{quantitative} 
treating of time, in \cite{AL94} Abadi and Lamport show how to introduce 
a distinguished state variable $now$ with a \kw{continuous} domain, representing 
the current time, so that the specification of temporal properties 
consists of formulas predicating explicitly on the values of $now$ 
in different states, thus describing its expected behavior with 
respect to the events taking place.

With reference to the resource manager example, to formally describe 
the behavior in case of a low-priority request an action $\lpr$ 
would be introduced, describing the untimed behavior of this 
request. An \emph{action} is a predicate about two states, whose 
values are denoted by unprimed and primed variables, for the current and 
next state, respectively. Therefore, the untimed behavior of 
an accepted low-priority request would simply be to change the 
value of the state of the resource (indicated by a variable $res$)
from free to occupied, as in the following definition.
\begin{equation*}
	 \mathsf{lpr} \ \triangleq \ res = \text{free}  \ \wedge\   res' = \text{occ}
\end{equation*}
Then, the timed behavior associated with this action would be 
specified by setting an upper bound on the time taken by the 
action, specifying that the action must happen within 2 time 
units whenever it is continuously enabled. Following the scheme 
in \cite{AL94}, a \emph{timer} would be defined by means of two formulas 
(which we do not report here for the sake of brevity: the interested 
reader can find them in \cite{AL94}). The first one defines predicate $\mathrm{MaxTime}(t)$,
which holds in all states whose timestamp (represented by the 
state variable $now$) is less than or equal the absolute time $t$.
The second formula defines predicate $\mathrm{VTimer}(t, A, \delta, v)$,
where $A$ is an action, $\delta$ is a delay, $v$ is 
the set of all variables, and $t$ is a state variable representing 
a timer. Then, $\mathrm{VTimer}(t, A, \delta, v)$ holds if 
and only if either action $A$ is not currently enabled and 
$t$ is $\infty$, or $A$ is enabled and $t$ is $now + \delta$ 
(and it will stay so until either $A$ occurs, or $A$ is disabled, 
see \cite[Sec.~3]{AL94} for further details).

Finally, the timed behavior of low-priority requests would be 
defined by the following action $\mathsf{lpr}^t$, where $T_\mathrm{gr}$ is a state variable representing 
the maximum time within which action $\lpr$ must occur.
  \begin{equation*}
	 \mathsf{lpr}^t \ \triangleq \ \mathsf{lpr} \ \wedge \ \mathrm{VTimer}(T_\mathrm{gr}, \mathsf{lpr}, 2, v) \ \wedge\ \mathrm{MaxTime}(T_\mathrm{gr})
  \end{equation*}
More precisely, the formula above states that after action $\lpr$ is enabled,
it must occur before time surpasses value $now + 2$.

It is interesting to discuss how TLA solves the problem of \kw{Zeno} 
behaviors. Zeno behaviors are possible because TLA formulas involving 
time are simply satisfied by behaviors where the variable $now$, 
being a regular state variable, does not change value. There 
are at least two mechanisms to ensure non-Zenoness. The first, 
simpler one introduces explicitly in the specification the requirement 
that time always advances, by the following formula $\mathrm{NZ}$. 
  \begin{equation*}
	 \mathrm{NZ} \ \triangleq \ \forall t \in \reals: \FLTL\!(now > t)
  \end{equation*}
  An alternative \emph{a posteriori} approach, which we do not discuss
  in detail, is based on a set of theorems provided in \cite{AL94} to
  infer the non-Zenoness of specifications written in a certain
  canonical form, after verifying some semantic constraints regarding
  the actions included in the specification.

It is worth noticing that also in TLA, like in other temporal 
logics discussed above, two consecutive states may refer to the 
same time instant, so that the logic departs from the notion 
of time inherited from classical physics and from traditional 
dynamical system theory. In every timed TLA specification, it 
is thus customary to explicitly introduce a formula that states 
the separation of time-advancing steps from ordinary program 
steps (see \cite{AL94} for further details). This 
approach is somewhat similar in spirit to that adopted in TTM/RTTL. which is presented in Section \ref{sec:duallanguage}.

\paragraph{Timed Propositional Temporal Logic.}
The TPTL logic by Alur and Henzinger represents a quite interesting 
example of how a careful choice of the operators provided by 
a temporal logic can make a great difference in terms of expressiveness, 
decidability, and complexity of the verification procedures. 
TPTL may be roughly described as a ``half-order'' logic, in that 
it is obtained from propositional \kw{linear} time logic by adding 
variables that refer to time, and allowing for a \emph{freeze quantification} 
operator: for a variable $x$, the freeze quantifier (denoted 
as $x.$) bounds the variable $x$ to the time when the sub-formula in 
the scope of the quantification is evaluated. One can think of 
it as the analogue, for logic languages, of clock resets in timed 
automata (see Section \ref{sec:synchronous}). The freeze quantifier is combined 
with the usual modal operators $\FLTL$ and $\GLTL$: if $\phi(x)$
is a formula in which variable $x$ occurs free, then formula $\FLTL x. \phi(x)$
asserts that there is some future instant, with some absolute 
time $k$, such that $\phi(k)$ will hold in that instant; 
similarly, $\GLTL x. \phi(x)$ asserts that $\phi(h)$
will hold in any future instant, $h$ being the absolute time 
of that instant.

The familiar property of the resource manager, that any low priority 
resource request is satisfied within 100 time units would be 
expressed in TPTL as follows. 
\begin{equation*}
  \GLTL x. \left( \mathsf{lpr} \Rightarrow \FLTL y. \left( \mathsf{occ} \wedge y < x + 100 \right) \right)
\end{equation*}

In \cite{AH94} the authors show that the logic is decidable over \kw{discrete} 
time, and define a doubly exponential \kw{decision procedure} for 
it; in \cite{AH92} they prove that adding ordinary first order quantification 
on variables representing the current time, or adding past operators 
to TPTL, would make the decision procedure of the resulting logic 
non-elementary. Therefore they argue that TPTL constitutes the 
``best'' combination of expressiveness and complexity for a temporal 
logic with \kw{metric} on time.

\subsubsection{Algebraic Formalisms} \label{sec:algebraic}
Algebraic formalisms are descriptive formal languages that focus 
on the \emph{axiomatic} and \emph{calculational} aspects of a specification. 
In other words, they are based on axioms that define how one 
can symbolically derive consequences of basic definitions \cite{Bae04,Bae03}. 
From a software engineering viewpoint, this means that the emphasis 
is on \emph{refinement} of specifications (which is formalized through 
some kind of algebraic \emph{morphism}).

In algebraic formalisms devoted to the description of concurrent 
activities, the basic behavior of a system is usually called \emph{process}. 
Hence, algebraic formalisms are often named with the term \emph{process 
algebras}. A process is completely described by a set of (abstract) 
events occurring in a certain order. Therefore, a process is 
also called a \emph{discrete event system}.

In order to describe concurrent and reactive systems, algebraic 
formalisms usually provide a notion of \emph{parallel composition} 
among different, concurrently executing, processes. Then, the 
semantics of the global system is fully defined by applications 
of the transformation axioms of the algebra on the various processes. 
Such a semantics --- given axiomatically as a set of transformations 
--- is usually called \emph{operational semantics}, not to be confused 
with operational formalisms (see Section \ref{sec:operational}).

\paragraph{Untimed Process Algebras.}
Historically, the first process algebraic approaches date back 
to the early work by Beki{\v c} \cite{Bek71} and to Milner's comprehensive 
work on the Calculus of Communicating Systems (CCS) formalism \cite{Mil80,Mil89}.
Basically, they aimed at extending the axiomatic 
semantics for sequential programs to concurrent processes. In 
this section, we focus on Communicating Sequential Processes 
(CSP), another popular process algebra, introduced by Hoare \cite{Hoa78,Hoa85} 
and subsequently developed into several formalisms. As usual, 
we refer the reader to \cite{BPS01} for a more detailed and comprehensive 
presentation of process algebras, and to the historical surveys \cite{Bae04,Bae03}.

\emph{Communicating Sequential Processes} are a process algebra 
based on the notion of \emph{communication} between processes. The 
basic process is defined by the sequences of events it can generate 
or accept; to this end the $\rightarrow$ operator is used, which denotes a 
sequence of two events that occur in order. Definitions are typically 
recursive, and infinite behaviors can consequently arise
However, a pre-defined process \textsl{SKIP} always terminates as soon as it
is executed. In the following examples we denote primitive events 
by lowercase letters, and processes by uppercase letters.

Processes can be unbounded in number, and parametric with respect 
to numeric parameters, which renders the formalism very expressive. 
We exploit this fact in formalizing the usual resource manager 
example (whose complete CSP specification is shown in Table \ref{tab:untimedCSP}) 
by allowing an unbounded number of pending high-priority requests, 
similarly to what we did with Petri nets in Section \ref{sec:petrinets}.

In CSP two \emph{choice} operators are available. One is \emph{external} 
choice, denoted by the box operator $\Box$; this is basically a 
choice where the process that is actually executed is determined 
by the first (prefix) event that is available in the environment. 
In the resource manager example, external choice is used to model 
the fact that a \textsl{FREE} process can stay idle for one transition 
(behaving as process $\mathsf{P_N}$), or accept a high-priority request or 
a low-priority one (behaving as processes $\mathsf{P_H}$ and $\mathsf{P_L}$, respectively). 
On the other hand, \emph{internal} choice, denoted by the $\sqcap$ operator, 
models a nondeterministic choice where the process chooses between 
one of two or more possible behaviors, independently of externally 
generated events. In the resource manager example, the system's 
process $\mathsf{WG}$ internally chooses whether to skip once or twice 
before granting the resource to a low-priority request. A special 
event, denoted by $\tau$, is used to give a semantics to internal 
choices: the $\tau$ event is considered invisible outside 
the process in which it occurs, and it leads to one of the possible 
internal choices.

Concurrently executing processes are modeled through the parallel
composition operator $\Vert$.\footnote{ $\mathsf{P}_1 {_A}\Vert{_B}
  \mathsf{P}_2$ denotes the parallel composition of processes
  $\mathsf{P}_1$ and $\mathsf{P}_2$ such that $\mathsf{P}_1$ only
  engages in events in $A$, $\mathsf{P}_2$ only engages in events in
  $B$, and they both synchronize on events in $A\cap B$.} In our
example, we represent the occupied resource by a parallel composition
of an $\mathsf{OCC}$ process and a counter $\mathsf{CNT}(k)$ counting
the number of pending high-priority requests. The former process turns
back to behaving as a \textsl{FREE} process as soon as there are no
more pending requests.  The latter, instead, reacts to release and
high-priority request events. In particular, it signals the number of
remaining enqueued processes by issuing the parametric event
$\mathsf{enqueued}!k$ (which is received by an $\mathsf{OCC}$ process,
as defined by the incoming event $\mathsf{enqueued}?k$ of
$\mathsf{OCC}$).

\begin{table}[tbh]
\begin{center}
  \begin{eqnarray*}
	 \mathsf{FREE}   &   =   &   \Box_{k \in \{\mathsf{H},\mathsf{L},\mathsf{N}\}} \mathsf{P}_{k}   \\
	 \mathsf{P}_{\mathsf{N}}  &  =  &  \textsl{SKIP} \longrightarrow \mathsf{FREE} \\  
	 \mathsf{P}_{\mathsf{H}}  &  =  &  \mathsf{hpr} \longrightarrow \mathsf{P}_\mathsf{O} \\
	 \mathsf{P}_\mathsf{O}    &  = &  \mathsf{OCC} \ _{\{\mathsf{enqueued}\}}\Vert_{\{\mathsf{enqueued}, \mathsf{rel}, \mathsf{hpr}\}} \  \mathsf{CNT}(0) \\
	 \mathsf{OCC}  &  =  &  \mathsf{enqueued}?0 \longrightarrow \mathsf{FREE} %
                           \quad \Box \quad \mathsf{enqueued}?k: \naturals_{> 0} \longrightarrow \mathsf{OCC} \\
	 \mathsf{CNT}(-1)  &  =  &  \textsl{SKIP} \\
	 \mathsf{CNT}(k)  &  =  &  \mathsf{rel} \longrightarrow \mathsf{DEQ}(k) %
                              \quad \Box \quad \mathsf{hpr} \longrightarrow \mathsf{CNT}(k+1) \\
	 \mathsf{DEQ}(k)  &  =  &  \mathsf{enqueued}!k \longrightarrow \mathsf{CNT}(k-1) \\
	 \mathsf{P}_{\mathsf{L}}  &  =  &  \mathsf{lpr} \longrightarrow \mathsf{WG} \\
	 \mathsf{WG}  &  =  &  \mathsf{WG}_1 \sqcap \mathsf{WG}_2 \\
	 \mathsf{WG}_1  &  =  &  \textsl{SKIP}\ ;\ \mathsf{P}_\mathsf{O} \\
	 \mathsf{WG}_2  &  =  &  \textsl{SKIP}\ ;\ \textsl{SKIP}\ ;\ \mathsf{P}_\mathsf{O}
  \end{eqnarray*}
\end{center}
\caption{The resource manager modeled through CSP.}
\label{tab:untimedCSP}
\end{table}

Let us now discuss the characteristics of the process algebraic 
models in general --- and CSP in particular --- with respect to 
the time modeling issues presented in Section \ref{sec:dimensions}. 

\begin{itemize}
\item Basic process algebras usually have no \kw{quantitative} notion of 
time, defining simply an \emph{ordering} among different events. 
In particular, time is typically \kw{discrete} \cite{Bae04}. Variants 
of this basic model have been proposed to introduce metric and/or 
dense time; we discuss them in the remainder of this section.

\item The presence of the silent transition $\tau$ is a way of modeling 
\kw{nondeterministic} behaviors; in particular, the nondeterministic internal 
choice operator $\sqcap$ is based on the $\tau$ event.

\item Even if process algebras include nondeterministic behaviors, 
their semantics is usually defined on \kw{linear} time models. 
There are two basic approaches to formalize the semantics of 
a process algebra: the \emph{operational} one has been briefly discussed 
above; for the \emph{denotational} one we refer the interested reader 
to \cite{Sch00}.

\item The parallel \kw{composition} operation is a fundamental primitive 
of process algebras. The semantics which is consequently adopted 
for concurrency is either based on \emph{interleaving} or it is \emph{truly asynchronous}.
Whenever interleaving concurrency is chosen, it 
is possible to represent a process by a set of classes of equivalent 
linear \emph{traces} (see the timed automata subsection of Section \ref{sec:synchronous}). 
Therefore, the semantics of the parallel composition operator 
can be expressed solely in terms of the other operators of the 
algebra; the rule that details how to do this is called \emph{expansion theorem} \cite{Bae04}.
On the contrary, whenever a truly asynchronous 
concurrency model is chosen no expansion theorem holds, and the 
semantics of the parallel composition operator is not reducible 
to that of the other operators.


\item Processes described by algebraic formalisms may include \kw{deadlocked} \emph{behaviors}
where the state does not advance as some process is blocked. 
Let us consider, for instance, the following process $\mathsf{P_i}$, 
which internally chooses whether to execute $\hpr \rightarrow \mathsf{P_i}$ or $\lpr \rightarrow \mathsf{P_i}$:
\begin{equation*}
  \mathsf{P_i} \quad  =   \quad  \mathsf{hpr} \longrightarrow \mathsf{P_i} \ \sqcap\  \mathsf{lpr} \longrightarrow \mathsf{P_i}
\end{equation*}
Process $\mathsf{P_i}$ may \emph{refuse} an $\lpr$ event offered by 
the environment, if it internally (i.e., independently of the 
environment) chooses to execute $\hpr \rightarrow \mathsf{P_i}$. In such a 
case, $\mathsf{P_i}$ would deadlock. It is therefore the designer's 
task to prove \emph{a posteriori} that a given CSP specification 
is deadlock-free.
\end{itemize}

Among other popular process algebras, let us just mention the 
Algebra of Communicating Processes (ACP) \cite{BW90} and other approaches 
based on the integration of data description into process formalization, 
the most widespread approach being probably that of LOTOS \cite{vEVD89,Bri89}.

\paragraph{Timed Process Algebras.}
Quantitative time modeling is typically introduced in process 
algebras according to the following general schema, presented 
and discussed by Nicollin and Sifakis in \cite{NS91}. First of all, 
each process is augmented with an \emph{ad hoc} variable that \kw{explicitly} 
represents time and can be \kw{continuous}. Time is global and all cooperating processes 
are synchronized on it. 

Then, each process's evolution consists of a sequence of two-phase 
steps. During the first phase, an arbitrarily long --- but \emph{finite} 
--- sequence of events occurs, while time does not change; basically, 
this evolution phase can be fully described by ordinary process 
algebraic means. During the second phase, instead, the time variable 
is incremented while all the other state variables stay unchanged, 
thus representing time progressing; all processes also \kw{synchronously} 
update their time variables by the same amount, which can possibly 
be infinite (divergent behavior).

Time in such a timestamp model is usually called \emph{abstract} 
to denote the fact that it does not correspond to concrete or 
physical time. Notice that several of the synchronous operational 
formalisms, e.g., those presented in Section \ref{sec:synchronous}, can also 
be described on the basis of such a time model. For instance, 
in synchronous abstract machines \emph{\`{a} la} Esterel \cite{BG92} the 
time-elapsing phase corresponds implicitly to one (discrete) 
time unit.

Assuming the general time model above, the syntax of process 
algebras is augmented with constructs allowing one to \kw{explicitly} 
refer to \kw{quantitative} time in the description of a system. This 
has been first pursued for CSP in \cite{RR88}, and has been subsequently 
extended to most other process algebras. We refer the reader 
to \cite{BM02,NS91,Bae03} --- among others --- for more references, 
while briefly focusing on Timed CSP (TCSP) in the following example.

\begin{example}[Timed CSP]
The CSP language has been modified \cite{DS95,Sch00} by extending a minimal set of operators to allow 
the user to refer to metric time. In our resource manager 
example (whose complete Timed CSP specification is shown in Table \ref{tab:timedCSP}), 
we only consider two metric constructs: the special process \textsl{WAIT}
and the so-called timed timeout $\rhd^t$.

The former is a quantitative version of the untimed \textsl{SKIP}: $\textsl{WAIT}\; t$
is a process which just delays for $t$ time units. We use this 
to model explicitly the acceptance of a low-priority request, 
which waits for two time units before occupying the resource 
(note that we modified the behavior with respect to the untimed 
case, by removing the nondeterminism in the waiting time).

The timed timeout $\rhd^t$ a modification of the untimed timeout
$\rhd$ (not presented in the previous CSP example). The semantics of a
formula $\mathsf{P} \rhd^t \mathsf{Q}$ is that of a process that
behaves as $\mathsf{P}$ if any of $\mathsf{P}$'s initial events occurs
within $t$ time units; otherwise, it behaves as $\mathsf{Q}$ after $t$
time units. In the resource manager example, we exploit this semantics
to prescribe that the resource cannot be occupied continuously for
longer than 100 time units: if no release ($\rel$) or high-priority
request ($\hpr$) events occur within 100 time units, the process
$\mathsf{CNT}(k)$ is timed out and the process $\mathsf{DEQ}$ is
forcefully executed.
\begin{table}[tbh]
\begin{center}
\begin{eqnarray*}
  \mathsf{FREE}   &   =   &   \Box_{k \in \{\mathsf{H},\mathsf{L},\mathsf{N}\}} \mathsf{P}_{k}   \\
  \mathsf{P}_{\mathsf{N}}  &  =  &  \textsl{SKIP} \longrightarrow \mathsf{FREE} \\  
  \mathsf{P}_{\mathsf{H}}  &  =  &  \mathsf{hpr} \longrightarrow \mathsf{P}_\mathsf{O} \\
  \mathsf{P}_\mathsf{O}    &  = &  \mathsf{OCC} \ _{\{\mathsf{enqueued}\}}\Vert_{\{\mathsf{enqueued}, \mathsf{rel}, \mathsf{hpr}\}} \  \mathsf{CNT}(0) \\
  \mathsf{OCC}  &  =  &  \mathsf{enqueued}?0 \longrightarrow \mathsf{FREE} %
                         \quad \Box \quad \mathsf{enqueued}?k: \naturals_{> 0} \longrightarrow \mathsf{OCC} \\
  \mathsf{CNT}(-1) &  =  &  \textsl{SKIP} \\
  \mathsf{CNT}(k)  &  =  &  \left( \mathsf{rel} \longrightarrow \mathsf{DEQ}(k) \ \Box \ \mathsf{hpr} \longrightarrow \mathsf{CNT}(k+1) \right)
                                      \rhd^{100}  \mathsf{DEQ}(k) \\
  \mathsf{DEQ}(k)  &  =  &  \mathsf{enqueued}!k \longrightarrow \mathsf{CNT}(k-1) \\
  \mathsf{P}_{\mathsf{L}}  &  =  &  \mathsf{lpr} \longrightarrow \textsl{WAIT}\;2 \longrightarrow \mathsf{P}_\mathsf{O}
\end{eqnarray*}
\end{center}
\caption{The resource manager modeled through Timed CSP.}
\label{tab:timedCSP}
\end{table}
\end{example}

Finally, it is worth discussing how TCSP deals with the problem 
of \kw{Zeno} behaviors. The original solution of TCSP (see \cite{DS95}) 
was to rule out Zeno processes \emph{a priori} by requiring that 
any two consecutive actions be separated by a fixed delay of $\delta$
time units, thus prohibiting simultaneity altogether. This solution 
has the advantage of being simple and of totally ruling out problems 
of Zenoness; on the other hand, it forcefully introduces a discretization 
in behavior description, and it yields complications and lack of uniformity 
in the algebra axioms. Therefore, subsequent TCSP models have 
abandoned this strong assumption by allowing for simultaneous 
events and arbitrarily short delays. Consequently, the non-Zenoness 
of any given TCSP specification must be checked explicitly \emph{a posteriori}.

Several \kw{analysis} and \kw{verification} techniques have been developed 
for, and adapted to, process algebraic formalisms. For instance, 
let us just mention the FDR2 refinement checker \cite{Ros97}, designed 
for CSP, and the LTSA toolset \cite{MK99} for the analysis of dual-language 
models combining process-algebraic descriptions with labeled 
transition systems.

\subsection{Dual Language Approaches} \label{sec:duallanguage}
The dual language approach, as stated in the introduction 
of Section \ref{sec:descriptive}, combines an operational formalism, useful for 
describing the system behavior in terms of states and transitions, 
with a descriptive notation suitable for specifying its properties. 
It provides a methodological support to the designer, in that 
it constitutes a unified framework for requirement specification, 
design, and verification. Although a dual language approach often 
provides methods and tools for \kw{verification} (e.g., for model 
checking), we point out that effectiveness or efficiency of verification 
procedures are not necessarily a direct consequence of the presence 
of two, heterogeneous notations (an operational and a descriptive 
one), but can derive from other factors, as the case of SPIN, 
discussed below, shows. In recent years a great number of frameworks 
to specify, design and verify critical, embedded, real-time systems 
have been proposed, which may be considered as applications of 
the dual language approach. As usual we limit ourselves to mention 
the most significant features of a few representative cases.

\subsubsection*{The TTM/RTTL Framework}
The work of Ostroff \cite{Ost89} is among the first ones addressing 
the problem of formal specification, design, and verification 
of real-time systems by pursuing a dual language approach. It 
proposes a framework based on Extended State Machines and Real-Time 
Temporal Logic (ESM/RTTL). In later works, ESM have been extended 
to Timed Transition Models (TTM) \cite{Ost90,Ost99}.

The operational part of the framework (TTM) associates transitions 
with lower and upper bounds, referred to the value of a global, 
\kw{discrete} time clock variable. We briefly discussed the time model 
introduced by this formalism in Section \ref{sec:synchronous}.

\begin{figure}[htb!]
	 \centering
	 \includegraphics{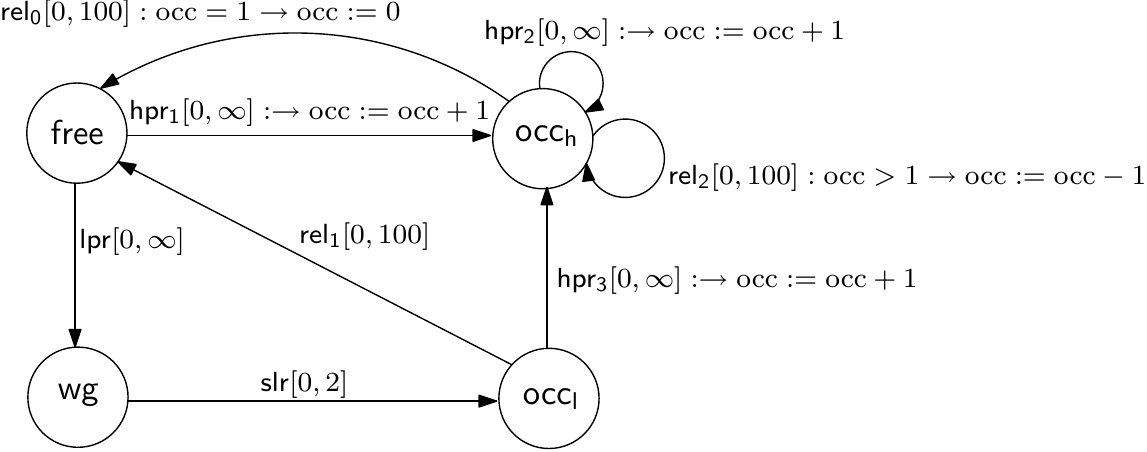}
	 \caption{A resource manager modeled through a Timed Transition Model.}
	 \label{fig:ttm}
\end{figure}

Here, let us illustrate TTM through the usual resource manager
example. Figure \ref{fig:ttm} represents a system similar to the Timed
Petri net example of Section \ref{sec:petrinets}: the number of
low-priority requests is not counted, while that of high-priority ones
is. Each transition is annotated with lower and upper bounds, a
\emph{guard}, and a variable update rule. For instance, the transition
$\rel_2$ can be taken whenever the guard $\occ > 1$ evaluates to true;
notice that we exploit an integer-valued state variable to count the
number of pending high-priority requests. The effect of $\rel_2$ is to
update the $\occ$ variable by incrementing it. Finally, when $\rel_2$
becomes enabled, it \emph{must} be taken within a maximum of 100 clock
ticks, \emph{unless} the state is left (and possibly re-entered) by
taking another (non tick) enabled transition (such as $\hpr_2$, which
is always enabled, since it has no guard).

The descriptive part of the TTM/RTTL framework (RTTL) is based 
on Manna and Pnueli's temporal logic: it assumes \kw{linear} time 
and it adopts the usual operators of future-only propositional 
LTL. Real-time (i.e., \kw{quantitative}) temporal properties are expressed 
by means of (in)equalities on simple arithmetic expressions involving 
the clock variable, as discussed in Section \ref{sec:temporallogics}. For instance, 
the familiar requirement that a low priority request is followed, 
within 100 time units, by the resource being occupied would be expressed 
as follows.
\begin{equation*}
 \forall \mathrm{T} \left( ( \lpr \wedge t = \mathrm{T}) \Rightarrow \FLTL ( \occ \wedge t \leq \mathrm{T} + 100 ) \right)
\end{equation*}

RTTL formulas are interpreted over TTM \emph{trajectories}, i.e., sequences of states
corresponding to TTM computations: \cite{Ost89} provides both a proof system and \kw{verification} procedures based 
on reachability analysis techniques.

The TTM/RTTL framework is also supported by the StateTime \kw{toolset} \cite{Ost97},
which in turn relies on the STeP tool \cite{BBCFMSU00}.

\subsubsection*{Model Checking Environments}
The SPIN model checking environment \cite{Hol03} is based, for the 
operational part, on B\"uchi automata, which are edited by the 
designer using a high-level notation called ProMeLa. The syntax 
of ProMeLa closely resembles that of the C programming language 
(and therefore is --- perhaps deceptively --- amenable to C programmers) 
and, in addition to the traditional constructs for sequential 
programming, provides features like parallel processes, communication 
channels, nondeterministic conditional instructions. The descriptive 
notation is plain future-only LTL, with the known limitations 
concerning the possibility to express complex properties and 
quantitative time constraints already pointed out in Section \ref{sec:temporallogics}. 
Model checking in SPIN is performed by translating the LTL formula 
expressing the required property into a B\"uchi automaton and 
then checking that the languages of the two automata (that obtained 
from the ProMeLa program and the one coming from the LTL formula) 
are disjoint. It is therefore apparent that the distinction between 
the operational and the descriptive parts is maintained only 
in the user interface for methodological purposes, and it blurs 
during verification.

UPPAAL \cite{LPY97} is another prominent framework supporting model-checking 
in a dual language approach. The operational part consists of 
a network of timed automata combined by the CCS parallel composition 
operator, and it provides both synchronous communication and 
asynchronous communication. The descriptive notation uses CTL 
in a restricted form, allowing only formulas of the kind $\ACTL \GLTL \phi$,
$\ACTL \FLTL \phi$, $\EXCTL \GLTL \phi$, $\EXCTL \FLTL \phi$, and $\ACTL \GLTL (\phi \Rightarrow \ACTL \FLTL \psi)$,
where $\phi$ and $\psi$ are ``local'' formulas, i.e., Boolean expressions 
over state predicates and integer variables, and clock constraints.

\subsubsection*{Other Dual Language Approaches}
Among the numerous other dual language frameworks \cite{JM94} we mention \linebreak \cite{FMM94},
which combines timed Petri nets and the TRIO temporal logic: it provides a systematic
procedure for translating any timed Petri net into a set of TRIO axioms that characterize its 
behavior, thus making it possible to derive required properties 
of the Petri net within the TRIO proof system.

\cite{FM02} introduces a real-time extension of the Object Constraint 
Language (OCL, \cite{WK99}), which is a logic language that allows 
users to state (and verify through model checking) properties of transitions 
of UML state diagrams (which, as mentioned in Section \ref{sec:synchronous}, 
are a variation of Harel's Statecharts), especially temporal ones.


\section{Conclusions} \label{sec:discussion}
In computer science, unlike other fields of science and engineering, 
the modeling of time is often restricted to the formalization 
and analysis of specific problems within particular application 
fields, if not entirely abstracted away. In this paper we have 
analyzed the historical and practical reasons of this fact; we 
have examined various categories under which formalisms to analyze 
timing aspects in computing can be classified; then we surveyed 
--- with no attempt at exhaustiveness, but with the goal of conceptual 
completeness --- many of such formalisms; in doing so we analyzed 
and compared them with respect to the above categories.

The result is a quite rich and somewhat intricate picture of 
different but often tightly connected models, certainly much 
more variegate than the way time modeling is usually faced in 
other fields of science and engineering. As in other cases, in 
this respect, too, computing science has much to learn from other, 
more established, cultural fields of engineering, but also the 
converse is true \cite{GM06b}.

Perhaps, the main lesson we can extract from our study is that 
\emph{despite the common understanding that time is a basic, unique 
conceptual entity, there are ``many notions of time'' in our reasoning; 
this is reflected in the adoption of different formal models 
when specifying and analyzing any type of system where timing 
behavior is of any concern}.

In some sense the above claim could be seen as an application 
of a principle of relativity to the abstractions required by 
modern --- heterogeneous --- system design. Whereas traditional engineering 
could comfortably deal with a unique abstract model of time as 
an independent ``variable'' flowing in an autonomous and immutable 
way to which all other system's variables had to be related, 
the advent of computing and communication technologies, with 
elaboration speeds that are comparable with the light's speed 
produced, and perhaps imposed, a fairly sharp departure from 
such a view:

\begin{itemize}
\item Often a different notion of time must be associated with
  different system's components. This may happen not only because the
  various components (possibly social organizations) are located in
  different places and their evolution may take place at a speed such
  that it is impossible to talk about ``system state at time $t$'',
  but also because the various components may have quite different
  nature --- typically, a controlled environment and a controller
  subsystem based on some computing device --- with quite different
  dynamics.

\item In particular, even inside the same computing device, it may 
be necessary to distinguish between an ``internal time'', defined 
and measured by device's clock, and an ``external time'', which 
is the time of the environment with which the computing apparatus 
must interact and synchronize. The consequence of this fact 
is that often, perhaps in a hidden way, two different notions 
of time coexist in the same model (for instance, the time defined 
by the sequence of events and the time defined by a more or less 
explicit variable --- a clock --- whose value may be recorded and assigned 
just like other program variables).

\item A different abstraction on time modeling may be useful depending 
on the type of properties one may wish to analyze: for instance, 
in some cases just the ordering of events matters, whereas in 
other cases a precise quantitative measure of the distance among 
them is needed. As a consequence many different mathematical 
approaches have been pursued to comply with the various modeling 
needs, the distinction between discrete and continuous time domains 
being only ``the tip of the iceberg'' of this issue.
\end{itemize}

Whether future evolutions will produce a better unification of 
the present state of the art or even more diversification and 
specialization in time modeling is an open and challenging question.



\newcommand{\etalchar}[1]{$^{#1}$}

\end{document}